\documentclass[aoas]{imsart}

\RequirePackage{amsthm,amsmath,amsfonts,amssymb}
\RequirePackage[authoryear]{natbib}
\RequirePackage[colorlinks,citecolor=blue,urlcolor=blue]{hyperref}
\RequirePackage{graphicx}

\usepackage{caption}
\usepackage{subcaption}
\usepackage{float}
\usepackage{multirow}
\usepackage{lscape}
\usepackage{calc}
\usepackage{array}
\usepackage{bbold}
\usepackage{comment}
\usepackage{xcolor}
\usepackage{nicematrix}
\usepackage{multirow}
\usepackage{diagbox}
\usepackage{tikz}
\usepackage{yhmath}
\usetikzlibrary{quotes,angles}
\usepackage{pgfplots}
\pgfplotsset{compat = newest}
\usepackage{fancyhdr}

\startlocaldefs

\newtheorem{proposition}{Proposition}[section]

\theoremstyle{definition}

\theoremstyle{remark}

\theoremstyle{remark}

\newcommand {\R}{\mathbb{R}}
\newcommand {\E}{\mathbb{E}}

\endlocaldefs

\begin{document}

\begin{frontmatter}
\title{Varying coefficients correlated velocity models in complex landscapes with boundaries applied to narwhal responses to noise exposure}
\runtitle{Movement models in complex landscapes}

\begin{aug}
\author[A]{\fnms{Alexandre}~\snm{Delporte}\ead[label=e1]{alexandre.delporte@univ-grenoble-alpes.fr}},
\author[B]{\fnms{Susanne}~\snm{Ditlevsen}\ead[label=e2]{susanne@math.ku.dk}}
\and
\author[A]{\fnms{Adeline}~\snm{Samson}\ead[label=e3]{adeline.leclercq-samson@univ-grenoble-alpes.fr}}
\address[A]{Laboratoire Jean Kuntzmann,
	Université Grenoble-Alpes\printead[presep={,\ }]{e1,e3}}

\address[B]{Department of Mathematical Sciences,
	University of Copenhagen \printead[presep={,\ }]{e2}}
\end{aug}

\begin{abstract}
Narwhals in the Arctic are increasingly exposed to human activities that can temporarily or permanently threaten their survival by modifying their behavior. We examine GPS data from a population of narwhals exposed to ship and seismic airgun noise during a controlled experiment in 2018 in the Scoresby Sound fjord system in Southeast Greenland. The fjord system has a complex shore line, restricting the behavioral response options for the narwhals to escape the threats. We propose a new continuous-time correlated velocity model with varying coefficients that includes an attractive and a repulsive term, which imply spatial constraints by adjusting the direction of the movement to the boundary of the domain. To assess the sound exposure effect we compare a baseline model for the movement before exposure to a response model for the movement during exposure. Our model, applied to the narwhal data, indicates that sound exposure can disturb motion up to a couple of tens of kilometers. Specifically, we found an increase in velocity and a decrease in persistence. We give estimates and confidence intervals of threshold distances to the sound source for medium and strong shifts in the whales motion.
\end{abstract}


\end{frontmatter}


\section{Introduction}


There is a large variety of sources of anthropogenic underwater noise in the Arctic region, including sonars, ice breakers, vessel traffic, drilling  and seismic airguns used for oil and gas exploration \citep{halliday_underwater_2020}.
A better understanding of the impact of these noises on marine life is critical for conservation policies. Marine mammals, whose vital functions highly depend on sound perception for communication and orientation, are particularly vulnerable. Besides causing physical harm, anthropogenic noise can also disturb their behavior and interrupt their natural foraging habits. However, there is still no clear criteria for measuring behavioral disturbances, due to the multiplicity of contextual variables that can influence a behavioral reaction, and the variety of these reactions \citep{southall_marine_2008,southall_marine_2019}. Comprehensive behavioral studies are thus essential.\\

This paper concentrates on the horizontal motion of narwhals. Previous analysis showed that the probabilities of the narwhals being close to the shore or moving towards the shore increase with exposure to the sound \citep{heide-jorgensen_behavioral_2021}. However, these analyses were based on discrete-time discrete-space Markov chains, despite the continuous nature of narwhals movement in both time and space. Especially space was reduced to three discrete states, reducing notably the amount of information contained in the data. The quantification of the exposure effect might thus be affected. \\

Animal movement is often modeled using continuous-time processes defined as solutions to stochastic differential equations (SDEs). One well-known example is the continuous-time correlated random walk, also called the integrated Ornstein-Uhlenbeck process or correlated velocity model (CVM) \citep{johnson_continuous_2008}. It can be viewed as a special case of velocity potential model, where the potential is a second degree polynomial \citep{preisler_analyzing_2013}.  Varieties of CVMs have been applied to birds \citep{janaswamy_state_2018}, marine algae \citep{gurarie_estimating_2011}, ants \citep{russell_spatially_2018}, seals \citep{johnson_continuous_2008,albertsen_generalizing_2018}, bowhead whales \citep{gurarie_correlated_2017} and sea lion motion \citep{hanks_reflected_2017}.\\

In our case study, the narwhals move in a restricted domain that consists in a complex system of fjords in Scoresby Sound (South-East Greenland), the largest fjord system in the world, known to be the summer residence for an isolated population of narwhals. Thus, the modeling process must include irregular shoreline boundaries through movement constraints. Most existing CVMs for animal movement do not account for such constraints \citep{johnson_continuous_2008,michelot_varying-coefficient_2021,gurarie_correlated_2017}.
Example models including landscape boundaries are constrained SDEs  \citep{hanks_reflected_2017} in which a reflected version of the CVM is considered to study sea lion telemetry data, and SDEs with drift described by a potential function \citep{russell_spatially_2018} where a repulsive potential function  constrains ants movement within a box. 
%
%
While these approaches effectively introduce movement constraints, they fail to capture certain behaviors. 
Narwhals have a proclivity to rotate to avoid or align with the shoreline without reaching the boundary. A reflected SDE as in \cite{hanks_reflected_2017} does not allow this type of behavior. In addition, a simple exponential potential function is suitable only with basic boundaries, such as the rectangular box in \cite{russell_spatially_2018}, but would break down for a complex boundary such as Scoresby Sound fjords. As a consequence, there is a need to develop new models adapted to such spatial constraints and behaviors. \\

In this paper, we extend the rotational CVM defined in \cite{gurarie_correlated_2017} by considering a rotation parameter $\omega$ expressed as a smooth function of the distance to the boundary and the angle between the animal's heading and the boundary normal vector with both an attractive and a repulsive term. This makes the drift in the velocity equation dependent on the location process and the boundary of the domain, and allows the velocity to rotate as the animal approaches the boundary of the domain. We show that parameter estimation for this SDE model can be performed in the framework of the \texttt{R} package \texttt{smoothSDE} \citep{michelot_varying-coefficient_2021}. To make this possible, we derive explicit formulas for the transition density of the location and velocity process and formulate an approximate linear Gaussian state-space model from the discretisation of these formulas. Maximum likelihood estimation based on the Kalman filter is then performed as for any other model in \texttt{smoothSDE}. The estimation procedure exploits the capabilities of the \texttt{R} package \texttt{TMB} to include random effects and get approximate likelihood using Laplace's approximation \citep{kristensen_tmb_2016,albertsen_fast_2015}. It is considerably less computationally intensive than an estimation procedure based on Markov Chain Monte Carlo posterior samples as done in \cite{hanks_reflected_2017,russell_spatially_2018}.\\

In this general framework, sound exposure is introduced in the model through smooth functions of a covariate defined as the inverse of the distance to the sound source. This choice is motivated by \cite{heide-jorgensen_behavioral_2021, tervo_narwhals_2021,tervo_stuck_2023}.  We estimate the effect of the exposure variable on both the velocity and the persistence parameters of the diffusion process.  A response model is compared to a baseline model for the narwhal movement before exposure to the disturbance, that is, under normal conditions \citep{michelot_continuous-time_2022}. The effect of sound exposure is assessed as a deviation in the response model parameters when compared to the baseline model parameters. 
We eventually give estimates and confidence intervals of threshold distances to the sound source for medium and strong shifts in the whales movement. These estimates can serve as a guideline for mitigation measures and conservation policies.  \\

To summarize, the main contributions of this paper are:
\begin{itemize}
	\item Definition of a constrained version of the CVM with attractive and repulsive terms, where deviation angles from shoreline and distance to shore are used to constrain the movement within a polygon, and align the velocity with the boundary of the domain.
	\item Derivation of explicit formulas for the transition density of the rotational CVM defined in \cite{gurarie_correlated_2017}, and addition of this model in the framework of  the \texttt{R} package \texttt{smoothSDE}, to enable the use of smooth parameters depending on external covariates, and estimation from noisy observations irregularly spaced in time.
	\item Statistical analysis of narwhal trajectories based on our SDE model, using the exposure covariate defined as the inverse of the distance to the sound source, and estimation of threshold distances from the sound source for strong and medium shift in the behaviour of the whales.
\end{itemize}

In Section 2, we give the general context of controlled exposure experiments as well as an overview of the narwhal data available for the analysis.
The diffusion models are then discussed in Section 3. Section 4 is dedicated to the statistical model used to infer the shore constraint and the sound exposure effects. Results of a simulation study are shown in Section 5. Finally, our models are applied to analyse the behavioral response of the narwhals in Section 6.

\section{Movement data of six narwhals in South-East Greenland}
\label{section: movement data}

\subsection{Motivation}
\label{subsection: controlled exposure experiments}

Several surveys of behavioral responses have been conducted on beaked whales, which have been regularly involved in stranding events following sonar exercises \citep{tyack_beaked_2011,cioffi_trade-offs_2022}. Studies have also been conducted on belugas \citep{martin_exposure_2023}, sperm whales \citep{madsen_quantitative_2006}, blue whales \citep{friedlaender_preymediated_2016}, and narwhals \citep{heide-jorgensen_behavioral_2021, tervo_narwhals_2021, tervo_stuck_2023}.
The most critical behavioral responses - those potentially altering a population's capacity to survive, reproduce or forage - include changes in movement speed and direction, avoidance reactions as well as modified dive profiles or vocalizations \citep{southall_marine_2008}.
Studies have shown that narwhals can exhibit some of these critical behavioral responses to sound exposure, such as avoidance reactions or changes of direction to move towards the shore \citep{heide-jorgensen_behavioral_2021}. A significant decrease in their buzzing rate has also been assessed as far as 40 km away from the sound source \citep{tervo_narwhals_2021}. These results are based on controlled exposure experiments conducted in 2017 and 2018 in a pristine area in South-East Greenland, still largely unaffected by noise pollution, and which is home to a declining population of narwhals \citep{garde_biological_2022}. Our study is based on these data collected in 2018.
 \\

\subsection{Description of the controlled exposure experiments}
\begin{figure}[h!]
\centering
\includegraphics[scale=0.6]{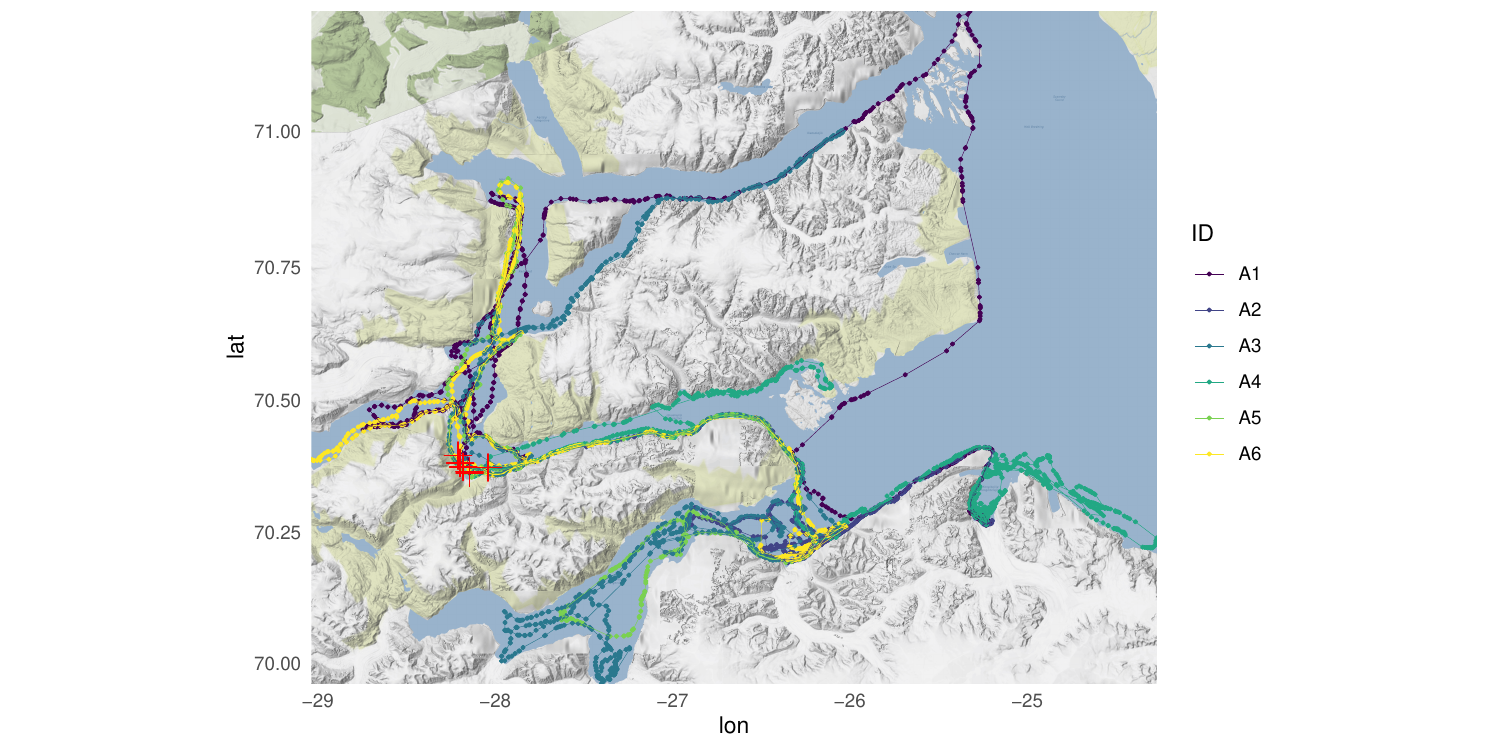}
\caption{Observed narwhals GPS tracks. Red crosses indicate the starting points of the trajectories.}
\label{fig: observed GPS data}
\end{figure}

We briefly recall the context for these controlled exposure experiments. For more details, the reader can refer to \cite{heide-jorgensen_behavioral_2021}. Six male narwhals were equipped with FastLoc GPS receivers in August 2018 in Scoresby Sound in South-East Greenland by biologists from the Greenland Institute of Natural Ressources, with the help of local Inuit hunters. 
An offshore patrol vessel military ship was sailed to shoot airguns underwater between August 25 and September 1. It was equipped with two airguns at 6m depth and moved at a speed of 4.5 knots. The guns were fired synchronously every $80$ s, while the GPS navigation system recorded the location of every shot. 
The data includes latitude and longitude of the narwhal positions, distance relative to the ship in metres, GPS time, and distance to the shore in metres. GPS positions are known only at times when the narwhals are at the surface. The median time step between two GPS measurements in the data is about $5$ minutes and only $0.3 \%$ of the time steps reach more than two hours, with a maximum at more than $4$ hours. While the statistical analysis in \cite{heide-jorgensen_behavioral_2021} relied on augmented data with positions linearly interpolated at each second, here we will only consider the actual GPS measurements with irregular time steps. The GPS tracks are shown in Figure~\ref{fig: observed GPS data}.\\

For each narwhal, the entire track is split into two periods: a period before exposure defined as the period before the narwhal gets in line of sight with the ship for the first time, and the period of exposure starting after exposure onset. We discard the GPS measurements from the first 12 hours after tagging, to avoid any tagging effects on the behavior \citep{nielsen_tagging2023}. We applied a velocity filter to keep only positions that have empirical velocity lower than 20 km/h. Two data points were removed with this filter.
Overall, $4815$ GPS positions are kept for the analysis. The splitting between data before and after exposure results in $1558$ measurements before exposure and $3257$ measurements after exposure.

\subsection{Notation and covariates}
\label{subsection: covariates}

For each narwhal $i \in \{1,\ldots,N\}$ with $N=6$, we denote $n_i$ the total number of observations,   decomposed as $n_i=n_{i,pre}+n_{i,post}$ where $n_{i,pre}$ is the number of observations before exposure, and $n_{i,post}$ is the number of observations after exposure onset. The positions are observed at discrete times $t_{i1}, t_{i2}, \cdots,t_{in_i}$ and  for $j \in \{1,\cdots,n_i\}$, we denote $y_{ij}=\begin{pmatrix} y_{ij1} & y_{ij2} \end{pmatrix}^\top$ the observed GPS position at time $t_{ij}$, projected in UTM zone 26 North coordinates with the \texttt{R} package \texttt{rgdal}. These points are noisy observations of the underlying unobserved true positions.\\

The land polygons define the boundaries of the domain $\mathcal{D}$ in which the narwhals can move. It is critical to accurately represent the shoreline geometry to ensure realistic movement constraints within the fjord system. Initially, we used land polygons obtained from OpenStreetMap, but these led to approximately $9\%$ of recorded narwhal positions being incorrectly classified as on land, likely due to inaccuracies in the shoreline representation. To address this, we applied a geometric buffer using the \texttt{st\_buffer} function in \texttt{R} to shrink the land polygons by 200 m. This adjustment significantly reduced the proportion of positions classified as on land to $0.12\%$ and yielded results more consistent with expected narwhal behavior. The shapefile defining the shoreline is made available in Supplementary materials. 

We then define two variables that will be included in the model, namely the distance to shore and the angle between the narwhal's heading and the shoreline. These variables are defined from smoothed trajectories of the observations. More precisely, denote $\tilde{y}_i$ the smooth trajectory for individual $i$ obtained by splines on a fine time grid. 
For each narwhal $i \in\{1, \ldots N\}$ and each time $t$, the closest point on the shoreline to the smoothed  position $\tilde{y}_{it}$  is denoted $p_{it}$. The distance $D^{shore}_{it}$ is defined as the distance between $\tilde{y}_{it}$ and $p_{it}$. It is an approximation of the actual distance to the boundary. Distance to shore values range from $0$ to $7.6$ km.
The angle $\Theta_{it}$ between the narwhal's heading and the shoreline is defined as the angle between the observed empirical velocity $\hat{v}_{it}=\frac{\tilde{y}_{it+\delta}-\tilde{y}_{it}}{\delta}$,  and the vector $\vec{n}_{it}=y_{it}-p_{it}$, as illustrated in Figure~\ref{fig: illustration nearest shore points}. A value $\Theta_{it}=\pm \frac{\pi}{2}$ indicates a movement parallel to the shore, while $\Theta_{it} \in \left]\frac{\pi}{2},\pi\right] \cup \left]-\pi,-\frac{\pi}{2}\right[$ indicates movement towards the shore.

\begin{figure}[ht!]
	\centering
	\begin{tikzpicture}
		\draw[gray, thick] (0,0) -- (0,4);
		\draw[red,fill=gray] (2,2) circle (.3ex);
		\draw[red,fill=red] (0,2) circle (.3ex);
		\draw[->,gray] (4,0) -- (2,2);
		\draw[->,gray,dashed,thick] (2,2) coordinate (O) -- (0,4) coordinate (A);
		\draw[->,red,dashed] (O) -- (4,2) coordinate (B);
		\draw[gray,dashed] (2,2) circle (2cm);
		\pic[draw=orange,->,angle eccentricity=1.2,angle radius=0.5cm] {angle=B--O--A};
		
		\node[left] at (-1,3.5) {Land};
		\node[right] at (1,3.5) {Water};
		\node[left] at (0,0) {$\partial \mathcal{D}$};
		\node[left] at (0,2) {$p_{it}$};
		\node[left] at (1.2,2.8) {$\hat{v}_{it}$};
		\node[below] at (1.8,2) {$\tilde{y}_{it}$};
		\node[below,red] at (3,2) {$\vec{n}_{it}$};
		\node[above,orange] at (2.3,2.3) {$\Theta_{it}$};
	\end{tikzpicture}
	\caption{Example of nearest point on the shore and angle $\Theta$. $\partial D$ represents the boundary of the domain, $\tilde{y}_{it}$ is the position of narwhal $i$ at time $t$.}
	\label{fig: illustration nearest shore points}
\end{figure}

\begin{figure}[ht!]
		\centering
		\includegraphics[scale=0.50]{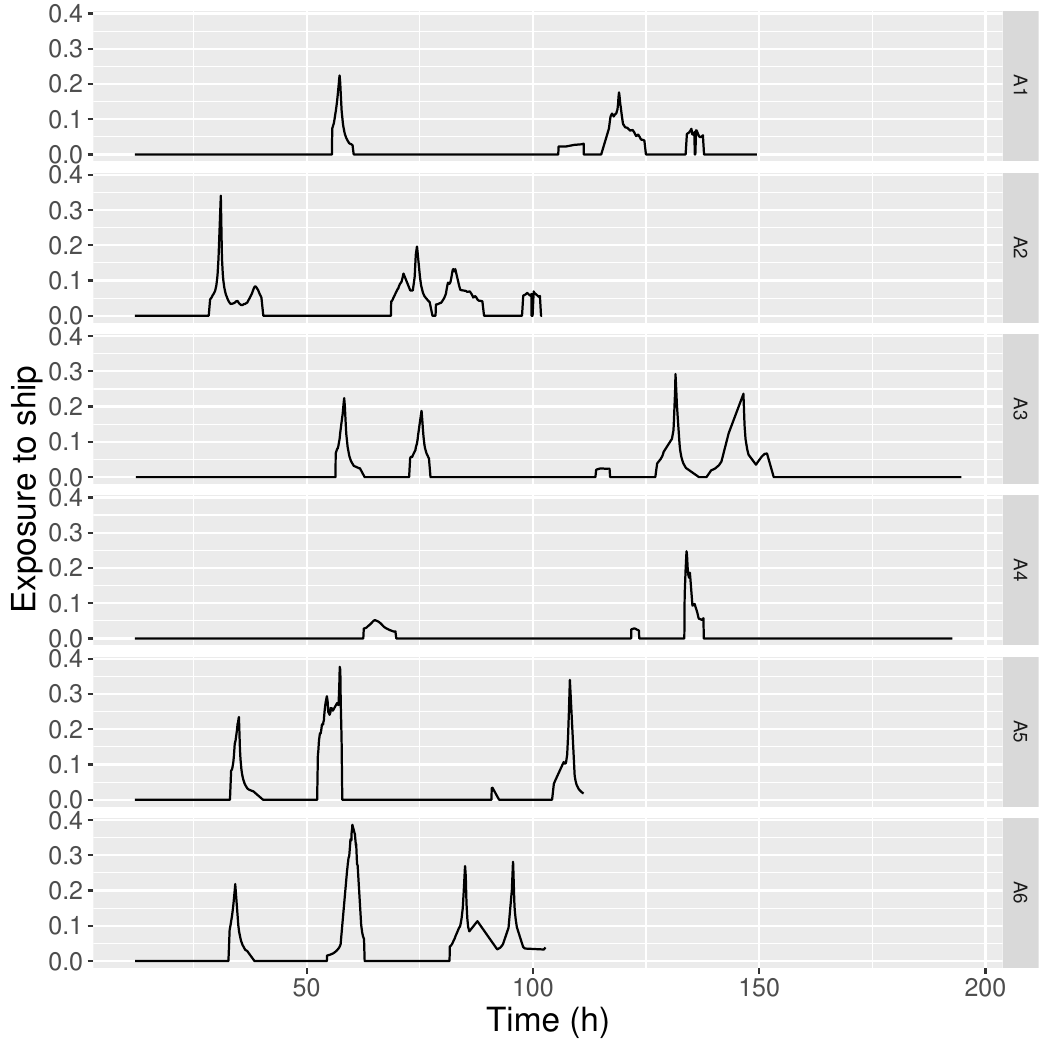}
	
	\caption{Covariate $E^{ship}$ (in $km^{-1}$) over time for each narwhal. Time 0 corresponds to 12 hours after tagging for each individual whale.}  
	\label{fig: realexpthroughtime}
\end{figure}

The distance to the ship $D^{ship}$ (km) is defined as the distance between observed GPS locations of the ship and the narwhal. The values are comprised between $2.7$ and $63.8$ km.
Exposure to the ship for narwhal $i$ at time $t_{ij}$, denoted $E^{ship}_{ij}$, is defined as the inverse of the distance to the ship (in km). The covariate $E^{ship}$ is meant to be a proxy for sound exposure levels received by the narwhals. Reasonably, the closer the narwhal is to the ship, the louder is the sound, and thus the exposure is larger. 
$D^{ship}$ is only defined when the narwhal is in line of sight with the ship and exposures are set to $0$ when this is not this case. This implies that in the statistical model, the ship noise is only allowed to affect the narwhal when the ship is in line of sight, though it is likely that narwhals can perceive the disturbance even when they are not in line of sight with the sound source. This provides a conservative estimate of the effect of the noise exposure. Levels of the covariate $E^{ship}$ for each narwhal is displayed in Figure~\ref{fig: realexpthroughtime}.

\section{Movement models within a constrained landscape}

The dynamics of the true positions of the narwhals is modelled by a SDE. First, we recall the standard Rotational Correlated Velocity Model (RCVM) \citep{gurarie_correlated_2017}, and then we extend it to include the effect of the distance to shore on the movement and constrain the position within a polygon.

\subsection{Standard Rotational  Correlated Velocity Model (RCVM)}
\label{section: RACVM}
\mbox{}\\

Let us recall the standard Rotational Correlated Velocity Model  proposed by \cite{gurarie_correlated_2017}. 
Let $A=\begin{pmatrix} 
	\frac{1}{\tau} &&-\omega \\
	\omega && \frac{1}{\tau}
\end{pmatrix}$ and define

\begin{equation} \left\{
	\begin{array}{l}
		dX(t)=V(t)dt \\
		dV(t)=-A(V(t)-\mu)dt+\frac{2\nu}{\sqrt{\pi \tau}} dW(t) 
	\end{array}
	\right.
	\label{eq: RACVM equation}
\end{equation}
where $V(t)$ is the horizontal two-dimensional velocity at time $t$, $X(t)$ is the two-dimensional horizontal position, typically in longitude and latitude or in UTM coordinates, $W(t)$ is a two-dimensional brownian motion. Parameter $\tau$ is an autocorrelation time scale, $\nu$ controls the norm of the velocity and drives the random variability, $\mu=\begin{pmatrix} \mu_1 && \mu_2 \end{pmatrix}^\top$ is the long term velocity, and $\omega$ is an angular velocity that controls how fast the velocity vector rotates. The case $\omega=0$ corresponds to the standard CVM without rotation defined by \cite{johnson_continuous_2008}. In most applications, this standard model is satisfactory, but in case of tortuous movement, a non-zero value of $\omega$ is sometimes needed \citep{gurarie_correlated_2017,alt_correlation_1990,albertsen_generalizing_2018}. We will later model $\omega$ as varying over time as a function of the distance to shore and the angle $\Theta$. Whether or not to include a non-zero mean velocity parameter $\mu$ depends on the specific context of the study. Typically, it makes sense to have $\mu\neq (0,0)$ when examining migratory patterns or avoidance reactions from a fixed sound source. We will later consider $\mu=0$ in our model.\\

In this section, we derive the explicit transition density for the process $U=\begin{pmatrix} X && V \end{pmatrix}^\top$. Though this model is considered in several papers, we found no mention of the closed form formulas derived here. In \cite{gurarie_correlated_2017}, only the process $V$ is exhaustively studied, and the classical distributional results about this process are used for estimation and simulation purposes.
A more general formulation is considered by \cite{albertsen_generalizing_2018} in which the diffusion matrix is lower triangular with positive diagonal elements and two distinct autocorrelation parameters are considered in each direction allowing for anisotropic movement. In this case, \cite{albertsen_generalizing_2018} derives the Gaussian transition density of the velocity process, where the covariance matrix is expressed using a Kronecker sum. However, the formulas for the process $X$
are not provided, and the Euler scheme is employed to approximate the transition densities of $X$. This approach is generally sufficient when the animal’s positions are recorded at high frequency. Yet, for marine mammals such as narwhals that dive to great depths, GPS measurements are usually taken at irregular intervals, typically of several minutes, which can render the Euler scheme unsuitable. Additionally, incorporating covariates into the parameters of the SDE often requires approximating the transition density by assuming the covariates remain constant during each time step \citep{michelot_varying-coefficient_2021}. This results in two layers of approximation, possibly leading to unreliable estimation.\\

Here, we derive a closed form formula for the transition density of the Markov process $(X ,V)$ under the hypothesis of isotropic movement, with diagonal diffusion matrix as in~\eqref{eq: RACVM equation}.\\

\begin{proposition}
	Let $U(t)=\begin{pmatrix} X(t) && V(t) \end{pmatrix}^\top$ for $t \geq 0$ be the solution to \eqref{eq: RACVM equation}. Then the exact transition density of the Markov process $U$ is given by 
	\begin{equation}
		U(t+\Delta) \vert U(t)=u \sim \mathcal{N}\left( T(\Delta) u +B(\Delta)\mu, Q(\Delta)\right) \mbox{ for all } \Delta>0 \mbox{ and } u \in \R^4
		\label{eq: X V distribution}
	\end{equation}
	where \begin{equation}
		T(\Delta)=\begin{pmatrix} I_2 && A^{-1}(I_2-e^{-A\Delta}) \\
			0_2 && e^{-A\Delta} \end{pmatrix} \mbox{, } B(\Delta)=\begin{pmatrix}
			\Delta I_2 -A^{-1}(I_2-e^{-A\Delta})\\
			I_2-e^{-A\Delta}\end{pmatrix}
		\label{eq: link matrices}
	\end{equation}
	and the covariance block matrix is given by
	\begin{equation}
		Q(\Delta)=\begin{pmatrix}
			q_1(\Delta)I_2 && \Gamma(\Delta) \\
			\Gamma(\Delta)^\top && q_2(\Delta)I_2
		\end{pmatrix}
		\label{eq: covariance matrix}
	\end{equation}
	with 
	\begin{equation*}q_1(\Delta)=\frac{\sigma^2}{C}\left( \Delta-2 \frac{\omega \sin(\omega \Delta)-\frac{1}{\tau} \cos(\omega \Delta)}{C} e^{ -\frac{\Delta}{\tau}} +\frac{\tau}{2} \left( \frac{\omega^2-\frac{3}{\tau^2}}{C}-e^{ -\frac{2\Delta}{\tau}}\right) \right)
	\end{equation*}
	
	\begin{equation*}
		q_2(\Delta)=\frac{2\nu^2}{\pi}\left(1-e^{-\frac{2 \Delta}{\tau}}\right), \quad
    \Gamma(\Delta)=\begin{pmatrix} \gamma_1 && \gamma_2 \\
			-\gamma_2 && \gamma_1\end{pmatrix}
	\end{equation*}
	where
	\begin{equation*}\gamma_1=\frac{\sigma^2}{2 C } \times \left( 1+e^{-\frac{2\Delta}{\tau}}-2e^{ -\frac{\Delta}{\tau}} \cos(\omega\Delta)\right)
	\end{equation*}
	\begin{equation*}\gamma_2=\frac{\sigma^2}{C} \times\left( e^{ -\frac{\Delta}{\tau}} \sin(\omega \Delta)-\frac{\omega \tau}{2} \left(1-e^{-2 \frac{\Delta}{\tau}} \right)\right)
	\end{equation*}
	and we denoted $C=\frac{1}{\tau^2}+\omega^2$ and $\sigma=\frac{2\nu}{\sqrt{\pi \tau}}$.
	\label{prop: transition density}
\end{proposition}

The formulas derived in \cite{johnson_continuous_2008} are a corollary of Proposition~\ref{prop: transition density}, obtained with $\omega=0$. The proof of this proposition is detailed in Appendix.

Equation~\eqref{eq: covariance matrix} shows that the two components of the location and velocity processes are independent.
Note that $e^{-A\Delta}=e^{\frac{-\Delta}{\tau}} \begin{pmatrix} \cos(\omega \Delta) && \sin(\omega \Delta) \\ -\sin(\omega \Delta) && \cos(\omega \Delta) \end{pmatrix}$ is a weighted rotation matrix of angle $-\omega \Delta$.
Intuitively, ~\eqref{eq: X V distribution} means that the velocity $V(t+\Delta)$ is a weighted mean of the long term mean velocity $\mu$ and the previous velocity $V(t)$ rotated by an angle $-\omega \Delta$.\\

For locomotion under spatial constraints, we hypothesize that an increase in tortuosity is a sign of avoiding the boundary and adapting the heading to the spatial constraint. We use this to define a constrained version of the standard RCVM.

\subsection{Constrained rotational correlated velocity model}
\label{section: CRCVM}

We propose a new RCVM that relies on the tortuosity parameter, or angular velocity $\omega$, to describe how the animals turn in reaction to the shore. 
To do so, we consider $\omega$ as a smooth function both of the distance to the shore $D^{shore}$ and the angle between the velocity vector and the shore normal vector $\Theta$. 
The process $X$ is constrained within a domain $\mathcal{D} \subset \R^2$ which is typically a polygon. The new model is 
\begin{equation} \left\{
	\begin{array}{l}
		dX(t)=V(t) dt \\
		dV(t)=-A(t)V(t)dt+\frac{2\nu}{\sqrt{\pi \tau}} dW(t) 
		
	\end{array}
	\right.
	\label{eq: CRCVM equation}
\end{equation}
with 
\begin{equation} 
	A(t)=\begin{pmatrix} 
		\frac{1}{\tau} && -\omega(t) \\
		\omega(t) && \frac{1}{\tau}
	\end{pmatrix}
	\label{eq: CRCVM matrix A}
\end{equation}
where $\omega(t)=f(\Theta(t),D^{shore}(t))$ with  $f$ a smooth function from $ ]-\pi,\pi] \times \R^{+}$ to $\R$. In the sequel, such models will be referred to as Constrained Rotational Correlated Velocity Models (CRCVM).\\
		

To induce a rotation when the animal is close to the boundary and heading in the direction of the shore,
the following assumptions must guide the shape of $f$:
\begin{itemize}
	\item[1)] for a distance to the shore lower than some threshold, $f(D^{shore},\cdot)$ should be positive and increasing on $]\frac{\pi}{2},\pi]$, 
	\item[2)] for a distance to shore  lower than some threshold, $f(D^{shore},\cdot)$ should be negative and decreasing on $]-\pi,-\frac{\pi}{2}[$,
	\item[3)] as distance to shore decreases, the magnitude of the functions  $f(D^{shore},\cdot)$ on $]\frac{\pi}{2},\pi[$  and $]-\pi,-\frac{\pi}{2}[$ should increase.
\end{itemize}

\begin{figure}[ht!]
	\centering
		\includegraphics[scale=0.4]{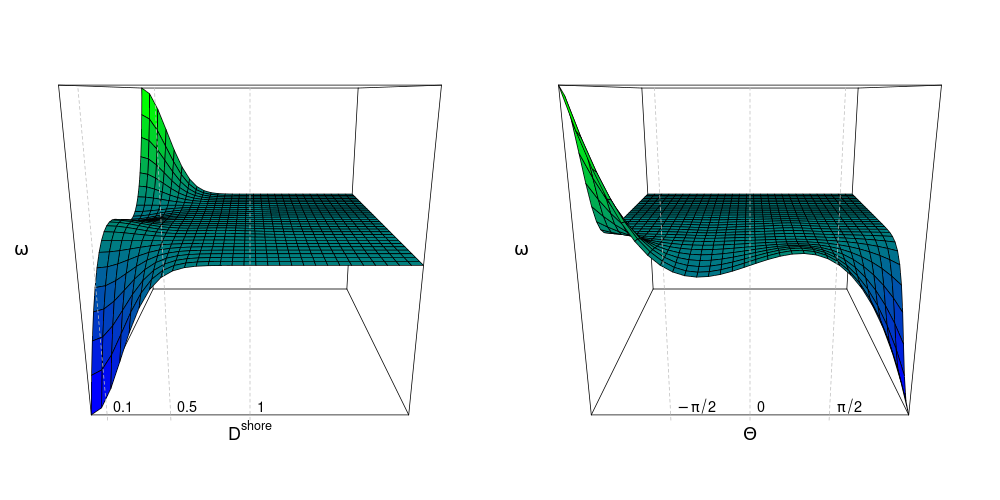}

\caption{Example of smooth function $f$ defined by~\eqref{eq: smooth omega}. Angular velocity $\omega$ increases in absolute value as $\Theta$ approaches $\pm\pi$, and decreases with increasing distance to the shore.  Here, angle $\Theta$ is in $rad$ and $D^{shore}$ in $km$. Parameter values are $a=\frac{10}{3}$, $b=5$, $D_r=0.3$ km,$D_a=0.8$ km, $\sigma_{\theta}=\pi/8$ rad , $\sigma_D=0.2$ km.}
\label{fig: perspectiveplots}
	
\end{figure}

We propose a parametric model for $f$. For $\Theta \in ]-\pi,\pi]$ and $D^{shore} > 0$, we define $f$ as the sum of a repulsive term and an attractive term:

\begin{equation}
	\label{eq: smooth omega}
	f(\Theta,D^{shore})=f_r(\Theta,D^{shore})+f_a(\Theta,D^{shore}))
\end{equation}
where the repulsive term is 
\[f_r(\Theta,D^{shore})=a\theta\left(\theta-\frac{\pi}{2}\right)\left(\theta+\frac{\pi}{2}\right)\times \frac{D_r}{D_{shore}}\exp\left(-\frac{D_{shore}}{D_r}\right)\]  and the attractive term is a Gaussian kernel
\[f_a(\Theta,D^{shore})=b\exp\left(-\left(\begin{pmatrix} \Theta & D^{shore} \end{pmatrix}-m\right) \Sigma^{-1} \left(\begin{pmatrix} \Theta & D^{shore}\end{pmatrix}-m)^\top \right)\right).\]
 Here, $ m=(\frac{\pi}{2\sqrt{3}},D_a)$, $\Sigma=\begin{pmatrix} \sigma_\Theta^2 & 0 \\ 0 & \sigma_D^2\end{pmatrix}$ and $D_r,D_a>0$, $a,b>0$ , $\sigma_D,\sigma_\Theta >0$ are parameters.
The parameter $a$ controls the magnitude of the angular velocity, $D_r$ controls the distance at which the velocity starts to rotate to model repulsion from the boundary and $D_a$ controls the distance at which the velocity starts to rotate to model attraction to the boundary. Then, $ \sigma_D$ and $\sigma_\theta$ are the standard deviations in the Gaussian kernel. Figure~\ref{fig: perspectiveplots}  shows function~\eqref{eq: smooth omega}.

Compared to the reflected CVM in \cite{hanks_reflected_2017}, model~\eqref{eq: CRCVM equation} is very flexible, and different species-specific behaviours can be described. The function $f$ controls at which distance the animal will turn away from the shore, and how fast it will turn.  

\begin{figure}[ht!]
	\includegraphics[scale=0.42]{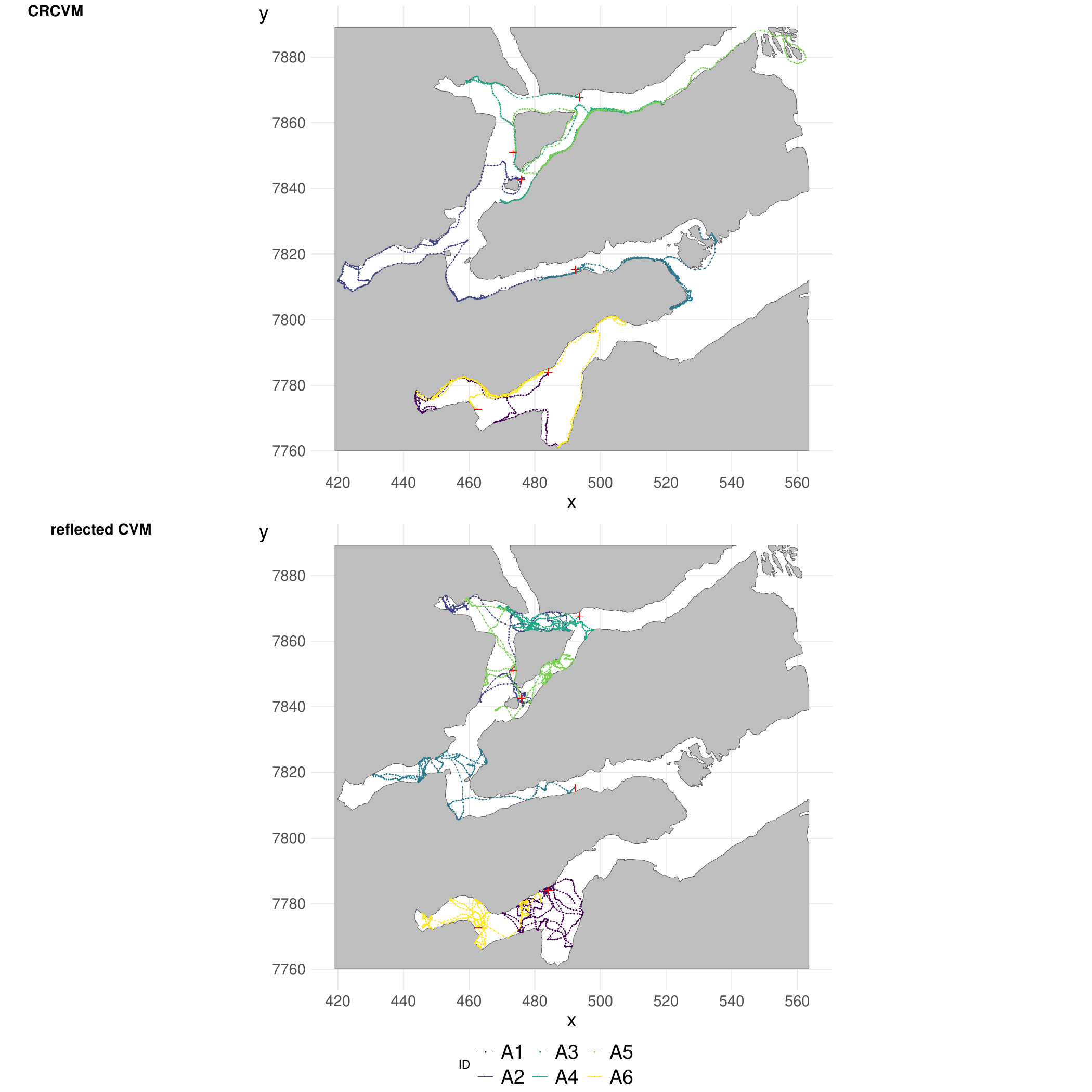}

	\caption{Three days long simulated trajectories for $6$ different initial positions within the fjords domain for the CRCVM defined by equation~\eqref{eq: CRCVM equation} (top) and the reflected CVM in \cite{hanks_reflected_2017} (bottom). Trajectories are simulated with fixed time step $\Delta= 10 $s and then subsampled to $10 $ min. All trajectories have constant velocity parameter $\nu=4$ km/h and constant persistence $\tau=2$ h.}
 \label{fig: crcvm examples}
\end{figure}

Solving explicitly~\eqref{eq: CRCVM equation} is out of reach due to the non-linearity induced by the distance to the shore and the angle $\Theta$. But it is possible to get an approximate solution. We choose a small time step $\Delta$ and approximate $D^{shore}$ and $\Theta$ by piecewise constant functions on interval $[k\Delta, (k+1)\Delta[$. We then use the formulas of Section~\ref{section: RACVM} to approximate the transition density of the model. 
Simulated trajectories obtained within the fjords domain are shown in Figure~\ref{fig: crcvm examples} and compared to trajectories simulated from a reflected CVM \citep{hanks_reflected_2017}. Trajectories of our model are effectively constrained and show different features from the reflected CVM: movement along and close to the boundary is favoured.   \\

In the following, the CRCVM \eqref{eq: CRCVM equation}, which allows $\omega$ to vary in time, is used to analyze narwhals trajectories. 

\section{Inference of the sound exposure effect}

To analyse potential perturbations in the locomotion of the narwhals due to sound exposure, we formulate a statistical mixed effect model for the parameters of the CRCVM, and discretise~\eqref{eq: CRCVM equation} to get an approximate state space model for inference with the Kalman filter.

\subsection{Mixed effect CRCVM for  shore influence and sound exposure}
\label{subsection: ship exposure effect}

First a baseline model is fitted only on the data before exposure, to get an estimate of natural behavior, when the narwhals are not exposed to the ship and airgun sounds. Then, a response model is fitted on the data during exposure to test whether the estimated smooth parameters deviate significantly from the baseline values. We present here the baseline and response models.\\

For the baseline model, only the effect of the shore is included through the covariates $D^{shore}$ and $\Theta$. For each narwhal $i \in \{1, \cdots, N\}$ and each time $j\in\{1, \cdots, n_{i,pre}\}$, $y_{ij}$ is an observation with measurement error of position $X_{ij}$:

\begin{equation}
	y_{ij}=X_{ij}+\varepsilon_{ij} \quad 
	\varepsilon_{ij} \underset{i.i.d}{\sim} \mathcal{N}(0,\sigma_{obs}^2)  
	\label{eq: baseline observations}
\end{equation}

The standard deviation $\sigma_{obs}$ represents GPS measurement errors variability. The assumption of i.i.d. Gaussian measurement errors is a simplified model of actual GPS measurement errors. In practice, the accuracy of GPS positions is influenced by the number of satellites processing the GPS signal—more satellites generally result in more accurate positions. Furthermore, the error can differ between the $x$ and $y$ directions. For a comprehensive study on Fastloc-GPS measurement errors, we refer to \cite{wensveen_path_2015}. It is often more appropriate to use a distribution with heavier tails, such as the Student's $t$-distribution, to model measurement error. However, this introduces additional complexity to the inference process, as the standard linear Kalman filter introduced by \cite{johnson_continuous_2008}
cannot be directly applied in such cases. Since the Student's $t$-distribution approaches a Gaussian distribution as its degrees of freedom increase, the Gaussian distribution is considered a reasonable trade-off between simplicity and suitability.\\

The latent processes $X_i$ and $V_i$ are solutions to

\begin{equation}  
	\left\{
	\begin{array}{l}
		dX_i(t)=V_i(t)dt  \\
		dV_i(t)=-\begin{pmatrix} 
			\frac{1}{\tau_i} && -\omega_i(t) \\
			\omega_i(t) && \frac{1}{\tau_i}
		\end{pmatrix}V_i(t)dt+\frac{2\nu_i}{\sqrt{\pi \tau_i}} dW(t) \\
	\end{array}
	\right.
	\label{eq:baseline latent processes}
\end{equation}
where $\tau_i$, $\nu_i$ are individual parameters   to account for variability between narwhals:
\begin{equation}   
	\begin{pmatrix}	\log(\tau_i)\\ \log(\nu_i) \end{pmatrix}
	= \begin{pmatrix}\log(\tau_0)\\ \log(\nu_0) \end{pmatrix}
	+
	\begin{pmatrix} b_{\tau,i}\\  b_{\nu,i}  \end{pmatrix}
	\quad \mbox{ with } \quad
	\begin{pmatrix} b_{\tau,i} \\ b_{\nu,i} \end{pmatrix} \underset{i.i.d}{\sim} \mathcal{N}\left(0,\begin{pmatrix} \sigma_{\tau}^2 && 0 \\ 0 && \sigma_{\nu}^2 \end{pmatrix}\right) 
	\label{eq:baseline random effects}
\end{equation}

Since $\tau$ and $\nu$ are positive, a log link function is used for these parameters. The coefficients $\log(\tau_{0})$, $\log(\nu_{0})$ are population intercepts and $b_{\tau,i}$, $b_{\nu,i}$
are the individual random effects. As discussed in Section \ref{section: CRCVM}, the angular velocity $\omega_i$ for individual $i$ is expressed as a smooth function of $\Theta$ and $D^{shore}$ : 
\begin{equation} 
	\omega_i(t)=f(D_i^{shore}(t),\Theta_i(t);a,b,D_a,D_r,\sigma_D,\sigma_\Theta)
\end{equation}
where the function $f$ is defined in equation \eqref{eq: smooth omega}. Due to identifiability issues on the real data, the parameters $a$, $b$,  $D_a$, $D_r$, $\sigma_D$ and $\sigma_\Theta$ will be considered as fixed and known.
\\

The response model is then designed to assess a deviation from the baseline model due to exposure to the ship by introducing the covariate $E^{ship}$ defined as the inverse of the distance to the ship. The observations are supposed to have the same measurement errors as the baseline. For $i \in \{1, \ldots, N\}$ and $j \in \{n_{i,pre}+1,\cdots,n_{i,post}\}$, we have

\begin{equation}  		y_{ij}=X_{ij}+\varepsilon_{ij}, \quad 		\varepsilon_{ij} \underset{i.i.d}{\sim} \mathcal{N}(0,\sigma_{obs}^2) 
	\label{eq: response observations}
\end{equation}

Then we add a dependency on the covariate $E^{ship}$ in the parameters $\tau_i$ and $\nu_i$: 
\begin{equation}    
	\begin{array}{l}
		\log(\tau_{i}(t))=\log(\tau_{0})+\alpha_{\tau} E^{ship}_i(t)+b_{\tau,i} \\
		\log(\nu_{i}(t))=\log(\nu_{0}) +  \alpha_{\nu} E^{ship}_i(t) +b_{\nu,i}  \\
	\end{array}
	\label{eq:response loglinear model}
\end{equation}

The intercepts $\log(\tau_0)$, $\log(\nu_0)$ and the smooth $\omega$ from the baseline model enter as offsets in the response model.
Thus, only the coefficients $\alpha_{\tau}$ and $\alpha_{\nu}$ are estimated from the data during exposure. These parameters control the deviation of $\tau_i$ and $\nu_i$ from their baseline values $\tau_0$ and $\nu_0$ as a function of the distance to the ship. Coefficients $\alpha_{\tau}$ and $\alpha_{\nu}$ significantly deviating from $0$ indicate an effect of the sound exposure on the narwhals horizontal motion.

\subsection{Approximate linear Gaussian state-space model}
\label{section: state space model}

In the baseline statistical model \eqref{eq:baseline latent processes}, we make the approximation suggested in \cite{michelot_varying-coefficient_2021} that the time-varying parameter $\omega_i$ is piecewise constant.
As explained in Section \ref{section: movement data}, on each time interval $[t_{ij},t_{ij+1}]$, we interpolate values of $D^{shore}_i$ and $\Theta_i$ at times $t_{ij}^{(0)} = t_{ij} < t_{ij}^{(1)} < \cdots < t_{ij}^{(k)}=t_{ij+1}$ with a time step $\Delta$ from the smoothed trajectoires $\tilde{y}_i$. 

We write $U_{ij}=\begin{pmatrix} X_{ij,1}  && X_{ij,2} && V_{ij,1} && V_{ij,2}\end{pmatrix}^\top$ and for $l \in \{0,\cdots,k\}$,
\[\omega_{ij}^{(l)}=\omega(t_{ij}^{(l)}), \quad   A_{ij}^{(l)}=\begin{pmatrix} 
	\frac{1}{\tau_{i}} && -\omega_{ij}^{(l)} \\
	\omega_{ij}^{(l)} && \frac{1}{\tau_{i}}
\end{pmatrix} \]

We use the formulas from Proposition \ref{prop: transition density} to obtain an approximation of the state-space matrix equations for each individual $i\in\{1,\ldots, N\}$, and each observation $j\in\{1, \ldots, n_{i,pre}\}$ when $\omega$ varies with time. We write $T_{ij}^{(l)}=T(\Delta_{ij}^{(l)})$ and $Q_{ij}^{(l)}=Q(\Delta_{ij}^{(l)})$ as defined in \eqref{eq: link matrices} and \eqref{eq: covariance matrix}.
Then the approximate transition density is 
	\begin{equation}
	U_{i,j+1} \vert U_{ij}=u \sim \mathcal{N}\left( \tilde{T}_{ij} u, \tilde{Q}_{ij}\right)
	\label{eq: approximate transition density}
\end{equation}
where $ \tilde{T}_{ij} = T_{ij}^{(k-1)} \times \cdots \times T_{ij}^{(0)}$ 
and $\tilde{Q}_{ij} =\sum_{l=1}^{k} T_{ij}^{(k-1)} \cdots T_{ij}^{(l+1)} Q_{ij}^{(l)} T_{ij}^{(l+1)\top} \cdots T_{ij}^{(k-1)\top}$. The calculations are detailed in Appendix.

Thus we obtain the following approximate linear state-space model 
\begin{equation}
	\begin{array}{l}
		y_{ij}=\begin{pmatrix} I_2 && 0_{2,2}\end{pmatrix}U_{ij}+\varepsilon_{ij},  \quad 
		\varepsilon_{ij} \underset{i.i.d}{\sim} \mathcal{N}(0,\sigma_{obs}^2) \\
		U_{i,j+1}=\tilde{T}_{ij}\,U_{ij} + \eta_{ij}, \quad 
		\eta_{ij} \sim \mathcal{N}(0,\tilde{Q}_{ij})
	\end{array}
	\label{eq: RACVM state space}
\end{equation}
\\
An approximate linear Gaussian state-space model is obtained similarly for the response model. Estimation with $\texttt{smoothSDE}$ relies on this state-space formulation.
The complete likelihood is computed from the observations $y_{ij}$ as a by-product of the Kalman filter algorithm \citep{michelot_varying-coefficient_2021}.
Laplace's approximation of the integral of this complete likelihood over the random effects is computed using the \texttt{R} package \texttt{TMB}  and optimization is performed via the \texttt{optim} function in $\texttt{R}$ with the BFGS gradient method, with the gradient being calculated by automatic differenciation. We refer to \cite{kristensen_tmb_2016} for more details about the derivation of the multidimensional Laplace's approximation and the gradient computations.
This method is already implemented for the CVM ($\omega=0$) in \texttt{smoothSDE}. Here, we extend this inference method to the more general model~\eqref{eq: CRCVM equation} and use it for estimation of the baseline and the response models.

\section{Simulation study for the baseline model}
\label{section: simulation constrained motion}
To evaluate the model and its estimation method, 
we simulate CRCVM trajectories over one day in the Scoresby Sound fjord system. We consider observations of a baseline model as described in Section~\ref{subsection: ship exposure effect}. We fix $\tau_0=1.5$ h and $\nu_0=4$ km/h. The random effects standard deviations are set to $\sigma_{\tau}=0.2$ and $\sigma_{\nu}=0.1$. The smooth function $\omega$ is defined as in equation \eqref{eq: smooth omega} with parameter values $D_r=0.5$, $D_a=1$, $a=4$, $b=4$, $\sigma_\theta=\frac{\pi}{8}$, $\sigma_D=0.3$. Initial velocities are set to $(0,0)$.  Initial positions are uniformly sampled between $0.1$ and $1$ km from the shore in the fjord system. Only $1.02\%$ of the simulated trajectories reached land. The batches containing at least one trajectory that reached land were not used for the estimation.\\

Different simulation designs are considered: high frequency data with either high or low measurement error, and low frequency data with high measurement error. For each setup, we simulate $M=100$ batches of $N$ individual trajectories with $N \in \{6,12\}$. The time steps between consecutive observations are set to $\Delta_{high}=1$ min for high frequency and $\Delta_{low}=5$ min for low frequency. The trajectories are simulated with time step $\Delta =1$ min and then subsampled. 
Measurement noise is Gaussian with standard deviation $\sigma_{obs,high}=40$ m and $\sigma_{obs,low}=10$ m for high and low measurement errors. They are supposed to be fixed and known in this simulation study.
  
For the estimation, we fix initial parameter values to $\tau_{0}=1$ h, $\nu_{0} = 1$ km/h, $\sigma_{\tau}=0.5$, $\sigma_{\nu}=0.5$. 
For each batch of $N$ trajectories, we obtain estimates $\widehat{\tau_{0}}^{(k)}$, $\widehat{\nu_{0}}^{(k)}$, $\hat{\sigma}_{\tau}^{(k)}$ and $\hat{\sigma}_{\nu}^{(k)}$, $k \in \{1,\cdots,M\}$ and compute the mean  and the $2.5\%$ and $97.5\%$ quantiles of the estimates. Simulation and optimization of the log-likelihood for one single batch of trajectories in all scenarios are performed on $16$ CPU cores in about $5$ h. Parallelization within \texttt{R} is used to simulate the trajectories, compute the angles $\Theta$ and the distances $D^{shore}$ and fit the models. 
\\

Parameters $(\tau_0, \nu_0, \sigma_\tau, \sigma_\nu)$ are estimated from the simulated data, considering known the parametric function $\omega$. 
This function depends on the covariates $D^{shore}$ and $\Theta$. We consider three versions of these covariates. First using the true ones, i.e. the ones calculated from the simulated $X(t)$ and $V(t)$, provides the best estimates that we could expect, even if they are not applicable with real data. Then we use $D^{shore}$ and $\Theta$ estimated from the smoothed trajectories $\tilde{y}_{i}$ as explained in Section \ref{section: movement data}. Trajectories are smoothed with the \texttt{gam} function in \texttt{R} software.  Third, we compare the gain of  smoothing the observations by  directly computing $D^{shore}$ and $\Theta$ on the raw measurements $y_{ij}$. Finally, we also apply the CVM model proposed by \cite{johnson_continuous_2008} to estimate the parameters without taking into account the domain barriers.

Estimated values are shown in Tables~\ref{table: simulation study hf parametric} and \ref{table: simulation study lf parametric} in the high and low frequency settings, respectively.
High frequency data allows to estimate precisely the parameters, except parameter $\tau_0$ which is the most difficult to estimate.  CVM gives the worst results. CRCVM with the ideal true covariates provides a good estimation of $\tau_0$. When the measurement error is low, the estimation obtained with the smoothed covariates is quite close, while the ones obtained from the observed covariates are biased. In the low frequency settings, the parameters $\nu_0$, $\sigma_\tau$ and $\sigma_\nu$ are well estimated. The best estimation of $\tau_0$ is obtained when using the smoothed interpolated covariates. 


	\begin{table}[ht!]
		\centering
		\begin{tabular}{|cccccc|}
			\hline
			     & &\multicolumn{2}{c}{\textbf{Intercepts}}  & \multicolumn{2}{c|}{\textbf{Random effects}} \\
			    \textbf{Model} & &$ \tau_0 $ (h) & $\nu_0 $ (km/h)& $\sigma_{\tau}$ & $\sigma_{\nu}$  \\
			&  & $1.50$ & $4.00$ & $0.2$ & $0.1$ \\
			\hline
			\multicolumn{6}{|c|}{\textbf{Low measurement error} ($\sigma_{obs} =10$ m)} \\
			\hline
            \multirow{4}{6em}{CVM} & \multirow{2}{4em}{$N=6$}  & $0.94$ &
			$4.09$ &
			$0.24$ & $0.13$ 
			\\
			& & $[0.73,1.10]$ &
			$[3.75,4.56]$ &
			$[0.16,0.41]$ & 
			$[0.07,0.23]$ 
			
			\\
			& \multirow{2}{4em}{$N=12$} & $0.92$ &
			$4.01$ &
			$0.26$ & $0.12$ 
			\\
			& & $[0.81,1.04]$  &
			$ [3.83,4.27]$ &
			$ [0.15,0.39]$ & $ [0.07,0.20]$ 
            \\
            \hline
			\multirow{4}{6em}{CRCVM \hspace{0.1cm} True} & \multirow{2}{4em}{$N=6$} & $1.42$ &
			$4.09$ &
			$0.23$ &
			$0.12$
			
			\\
			& &	$[1.20,1.74]$ &
			$ [3.77,4.49] $&
			$[0.17,0.28]$ & $ [0.09,0.28]$ 
			\\
			& \multirow{2}{4em}{$N=12$} & $1.48$ &
			$4.03$ & $0.25$ & $0.10$
			
			\\
			& & $[1.29,1.72]$ &
			$  [3.79,4.33]$ &
			$ [0.18,0.30]$ &
			$ [0.08,0.17]$ 
			\\
            \hline
            \multirow{4}{6em}{CRCVM Observed} & \multirow{2}{4em}{$N=6$} & $1.16$ &
			$4.12$ &
			$0.24$ &
			$0.13$
			
			\\
			& &	$[0.93,1.44]$ &
			$[3.76,4.50]$&
			$[0.16,0.36]$ & $[0.07,0.30]$ 
			\\
			& \multirow{2}{4em}{$N=12$} & $1.15$ &
			$4.01$ & $0.25$ & $0.10$
			
			\\
			& & $ [1.01,1.30]$ &
			$ [3.75,4.25]$ &
			$[0.17,0.34]$ &
			$[0.07,0.16]$ 
			\\
			\hline	
			\multirow{4}{6em}{CRCVM Smoothed} & \multirow{2}{4em}{$N=6$}  & $1.27$ &
			$4.10 $ &
			$0.24$ & $0.14$ 
			\\
			& & $[1.02,1.51]$ &
			$[3.74,4.47]$ &
			$[0.15,0.36]$ & 
			$[0.08,0.40]$ 
			
			\\
			& \multirow{2}{4em}{$N=12$} & $1.29$ &
			$4.03$ &
			$0.28$ & $0.12$ 

			\\
			& & $[1.11,1.42]$  &
			$ [3.85,4.20]$ &
			$ [0.18,0.39]$& $[0.07,0.18] $  \\
            \hline
			\multicolumn{6}{|c|}{\textbf{High measurement error} ($\sigma_{obs} =40$ m)} \\
			\hline
            \multirow{4}{6em}{CVM} & \multirow{2}{4em}{$N=6$}  & $0.92$ &
			$4.06$ &
			$0.25$ & $0.13$ 
			\\
			& & $[0.76,1.08]$ &
			$[3.74,4.44]$ &
			$[0.17,0.38]$ & 
			$[0.08,0.29]$ 
			\\
			& \multirow{2}{4em}{$N=12$} & $0.89$ &
			$4.00$ &
			$0.24$ & $0.12$ 
			\\
			& & $[0.79,1.02]$  &
			$ [3.84,4.26]$ &
			$ [0.16,.038]$ & $ [0.07,0.18]$ 
            \\
            \hline
            \multirow{4}{6em}{CRCVM \hspace{0.1cm} True} & \multirow{2}{4em}{$N=6$} & $1.40$ &
			$4.07$ &
			$0.24$ &
			$0.12$
			\\
			& &	$[1.15,1.76]$ &
			$ [3.71,4.49] $&
			$[0.17,0.60]$ & $ [0.08,0.34]$ 
			\\
			& \multirow{2}{4em}{$N=12$} & $1.43$ &
			$3.98$ & $0.27$ & $0.10$
			
			\\
			& & $[1.27,1.60]$ &
			$  [3.78,4.20]$ &
			$ [0.19,0.33]$ &
			$ [0.07,0.17]$ 
			\\
            \hline
            \multirow{4}{6em}{CRCVM Observed} & \multirow{2}{4em}{$N=6$} & $0.41$ &
			$4.14$ &
			$0.46$ &
			$0.13$
			
			\\
			& &	$[0.27,0.61]$ &
			$[3.71,4.57]$&
			$[0.17,0.75]$ & $[0.06,0.24]$ 
			\\
			& \multirow{2}{4em}{$N=12$} & $0.40$ &
			$4.06$ & $0.46$ & $0.13$
			
			\\
			& & $ [0.30,0.48]$ &
			$ [3.72,4.28]$ &
			$[0.28,0.64]$ &
			$[0.07,0.21]$ 
			\\
			\hline	
			\multirow{4}{6em}{CRCVM Smoothed} & \multirow{2}{4em}{$N=6$}  & $1.21$ &
			$4.07 $ &
			$0.24$ & $0.14$ 
			\\
			& & $[1.00,1.42]$ &
			$[3.71,4.45]$ &
			$[0.17,0.36]$ & 
			$[0.08,0.40]$ 
			
			\\
			& \multirow{2}{4em}{$N=12$} & $1.23$ &
			$4.01$ &
			$0.27$ & $0.11$ 

			\\
			& & $[1.09,1.43]$  &
			$ [3.82,4.21]$ &
			$ [0.19,0.36]$& $[0.07,0.18] $  \\
            \hline

		\end{tabular}
        \vspace{0.1cm}
		\caption{Results for the simulation study in the high frequency settings. Numbers are average and $95 \%$ CI of the estimates obtained from $M=100$ simulated data sets with  $N=6$ or $12$ individuals, low  (top) and high (bottom) measurement errors. The true values used in the simulations are given in the upper row. Four models are compared: CVM model without spatial constraints, CRCVM with the true covariates $D^{shore}$ and $\Theta$, with covariates calculated from observations $y$ or from their smoothed version $\tilde{y}$. }
		\label{table: simulation study hf parametric}
	\end{table}


\begin{table}[ht!]
		\centering
		\begin{tabular}{|cccccc|}
			\hline
			     & &\multicolumn{2}{c}{\textbf{Intercepts}}  & \multicolumn{2}{c|}{\textbf{Random effects}} \\
			
			     \textbf{Model} & &$ \tau_0 $ (h) & $\nu_0 $ (km/h)& $\sigma_{\tau}$ & $\sigma_{\nu}$  \\
		
			& & $1.50$ & $4.00$ & $0.2$ & $0.1$ \\
            \hline
            \multirow{4}{6em}{CVM} & \multirow{2}{4em}{$N=6$}  & $0.92$ &
			$4.08$ &
			$0.27$ & $0.14$ 
			\\
			& & $[0.73,1.12]$ &
			$[3.73,4.48]$ &
			$[0.19,0
            .42]$ & 
			$[0.08,0.32]$ 
			\\
			& \multirow{2}{4em}{$N=12$} & $0.91$ &
			$4.02$ &
			$0.25$ & $0.12$ 
			\\
			& & $[0.80,1.07]$  &
			$ [0.15,0.37]$ &
			$ [0.07,0.17] $ & $ [00.8,0.17]$ 
            \\
            \hline
            \multirow{4}{6em}{CRCVM \hspace{0.1cm} True} & \multirow{2}{4em}{$N=6$} & $0.90$ &
			$4.12$ &
			$0.31$ &
			$0.14$
			\\
			& &	$[0.70,1.14]$ &
			$ [3.76,4.53] $&
			$[0.19,0.48]$ & $ [0.09,0.23]$ 
			\\
			& \multirow{2}{4em}{$N=12$} & $0.90$ &
			$4.03$ & $0.31$ & $0.12$
			
			\\
			& & $[0.75,1.10]$ &
			$  [3.78,4.27]$ &
			$ [0.18,0.44]$ &
			$ [0.08,0.20]$ 
			\\
            \hline
            \multirow{4}{6em}{CRCVM Observed} & \multirow{2}{4em}{$N=6$} & $0.55$ &
			$4.20$ &
			$0.37$ &
			$0.15$
			
			\\
			& &	$[0.39,0.72]$ &
			$[3.77,4.68]$&
			$[0.16,0.68]$ & $[0.07,0.25]$ 
			\\
			& \multirow{2}{4em}{$N=12$} & $0.57$ &
			$4.09$ & $0.36$ & $0.14$
			
			\\
			& & $ [0.45,0.70]$ &
			$ [3.76,4.35]$ &
			$[0.18,0.58]$ &
			$[0.06,0.22]$ 
			\\
			\hline	
			\multirow{4}{6em}{CRCVM Smoothed} & \multirow{2}{4em}{$N=6$}  & $1.14$ &
			$4.12 $ &
			$0.27$ & $0.13$ 
			\\
			& & $[0.87,1.43]$ &
			$[3.76,4.61]$ &
			$[0.17,0.47]$ & 
			$[0.08,0.21]$ 
			
			\\
			& \multirow{2}{4em}{$N=12$} & $1.14$ &
			$4.00$ &
			$0.26$ & $0.12$ 

			\\
			& & $[0.97,1.36]$  &
			$ [3.73,4.27]$ &
			$ [0.17,0.39]$& $[0.07,0.20] $  \\
            \hline

		\end{tabular}
        \vspace{0.1cm}
		\caption{
        Results for the simulation study in the low frequency settings. Numbers are average and $95 \%$ quantiles of the estimates obtained from $M=100$ simulated data sets with  $N=6$ or $12$ individuals, and   high measurement errors. The true values used in the simulations are given in the upper row. Four models are compared: RCVM model without constraints, CRCVM with the true covariates $D^{shore}$ and $\Theta$, with covariates calculated from observations $y$ or from their smoothed version $\tilde{y}$.} 
    
		\label{table: simulation study lf parametric}
	\end{table}

\section{Application to narwhal data}
\label{section: application}

In this section, we apply our CRCVM to analyse the  behavioral response of the narwhals to ship and seismic airgun exposure.
The baseline model is fitted on the tracks before exposure to capture normal behavior. Deviations from the baseline is then assessed by fitting a model with covariate $E^{ship}$ on the tracks after exposure with the offsets estimated at baseline. Confidence intervals of the exposure effects are provided. Then the baseline model is numerically validated.

\subsection{Baseline estimation}
We estimate the baseline model described in Section~\ref{subsection: ship exposure effect} on the data before exposure. We initialize the random effects standard deviations to $\sigma_{\nu}=\sigma_{\tau}=0.5$, and fix the parameters for the  function $\omega$  to $D_r= 1.7 $ km, $D_a = 0.65$ km, which are respectively the $80 \%$ and $50\%$ quantiles of the observed distance to shore, $a=0.3$, $b=2$, $\sigma_D=0.5$ and $\sigma_\Theta =\frac{\pi}{6}$. 
The measurement error standard deviation is also fixed to $\sigma_{obs}=50$ m based on the measurement errors of Fastloc-GPS found in \cite{dujon_accuracy_2014,wensveen_path_2015}.  \\

Estimation and confidence intervals are shown in Table~\ref{table: narwhal data estimations}. 
The population mean for the persistence parameter is estimated to $\hat{\tau}_{0}=1.13 $ h, with $95\%$ confidence interval $(0.91; 1.40)$.
In comparison, harbour seal in Alaska were shown to exhibit slightly more persistent motion $\hat{\tau}=1.51$ h $(1.30 -1.75)$  \citep{johnson_continuous_2008} while bowhead whales in Greenland showed much less persistence $\hat{\tau}=0.17$ h  $(0.14 -0.20)$  \citep{gurarie_correlated_2017}.
The population mean for the velocity parameter is estimated to $\hat{\nu}_{0}=4.65$ km/h ($4.14; 5.21$) which is of the order of magnitude of the average of the observed velocities.

\subsection{Response estimation}

The baseline estimates of the intercepts $\log(\tau_0)$, $\log(\nu_0)$ and the standard deviations $\sigma_{\tau}$, $\sigma_{\nu}$ are used as offsets in the response model. Only the sound exposure effects on $\tau$ and $\nu$ are estimated from the data after exposure. The uncertainty stemming from the baseline estimates is considered by simulating  $100$ posterior samples of the baseline estimates using the Hessian matrix of the log-likelihood to approximate the covariance, and using the sampled parameters for the estimations of the response model. The mean of the estimates and the $95\%$ confidence intervals obtained over the $100$ estimations are shown in Table~\ref{table: narwhal data estimations}. As the coefficients $\alpha_\tau$ and $\alpha_\nu$ significantly deviate from zero, this suggests that sound exposure influences both the persistence and the velocity of movement.\\

\begin{table}[ht!]
	\centering
	\begin{tabular}{|lcc|}
		\hline
		Parameter & Estimate & CI \\ 
		\hline
        \multicolumn{3}{|l|}{\textbf{Baseline}} \\
        \hline
		$\tau_0 $ (h) & $1.13$ & $[0.91; 1.40]$ \\ 
		$\nu_0$ (km/h)& $4.65$ & $[4.14; 5.21]$ \\ 
		$\sigma_{\tau}$ (h) & $0.21$ & $[0.03; 0.73]$ \\ 
		$\sigma_{\nu}$ (km/h) & $0.11$ & $[0.04; 0.26]$  \\ 
		\hline
        \multicolumn{3}{|l|}{\textbf{Response}} \\
        \hline
        $\alpha_{\tau}$ & $-3.24$ & $[-4.67; -1.70]$ \\ 
        $\alpha_{\nu}$ & $1.10$ & $[0.49; 1.68]$ \\ 
        \hline
	\end{tabular}
	\caption{Results for real baseline data and real exposure data with CRCVM and smoothed covariates, with a log-linear exposure model}
\label{table: narwhal data estimations}
\end{table}

\begin{figure}[ht!]
	\begin{subfigure}{0.98\textwidth}\centering
		\includegraphics[width=0.7\linewidth]{"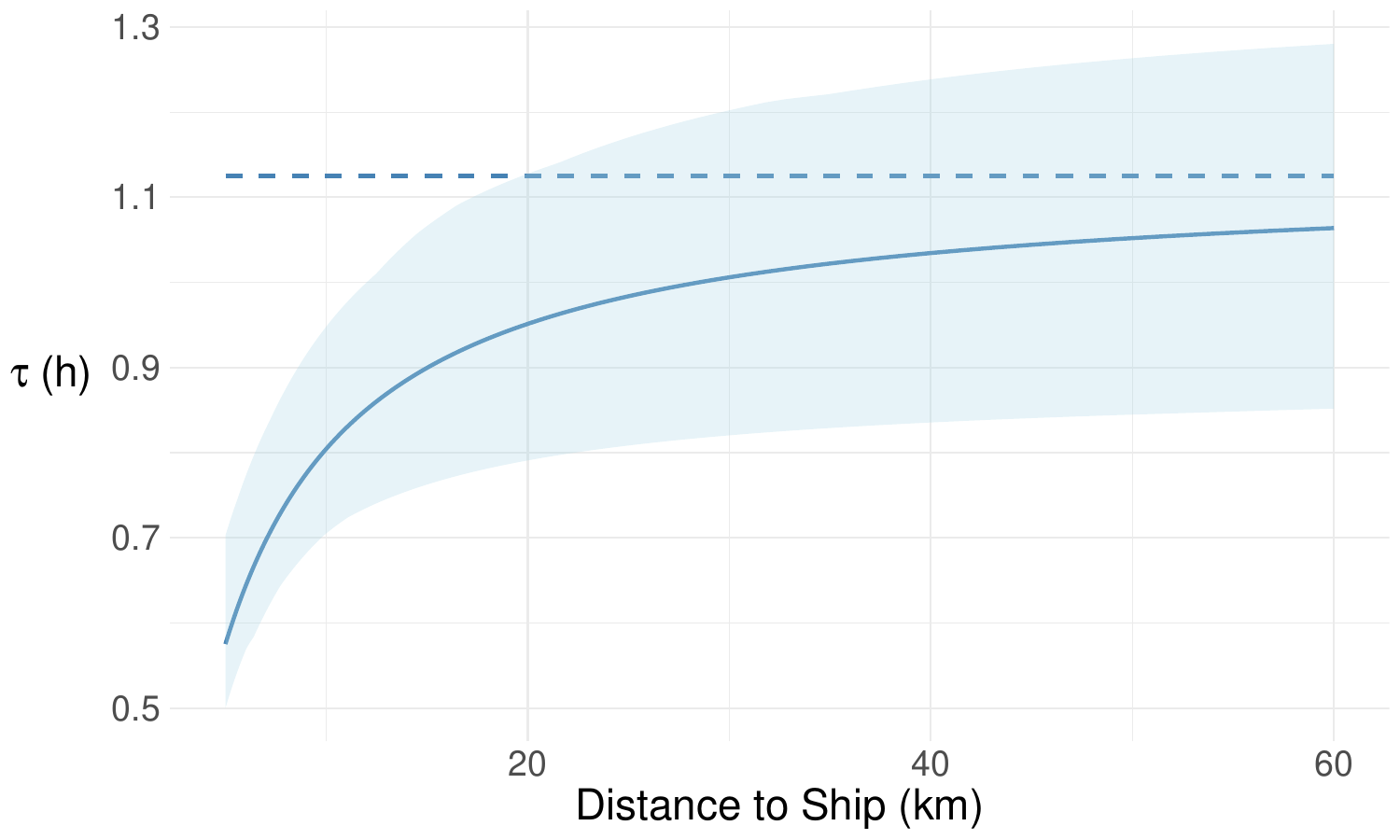"}
		\caption{}
	\end{subfigure}
	
	\begin{subfigure}{0.98\textwidth}
		\centering
		\includegraphics[width=0.7\linewidth]{"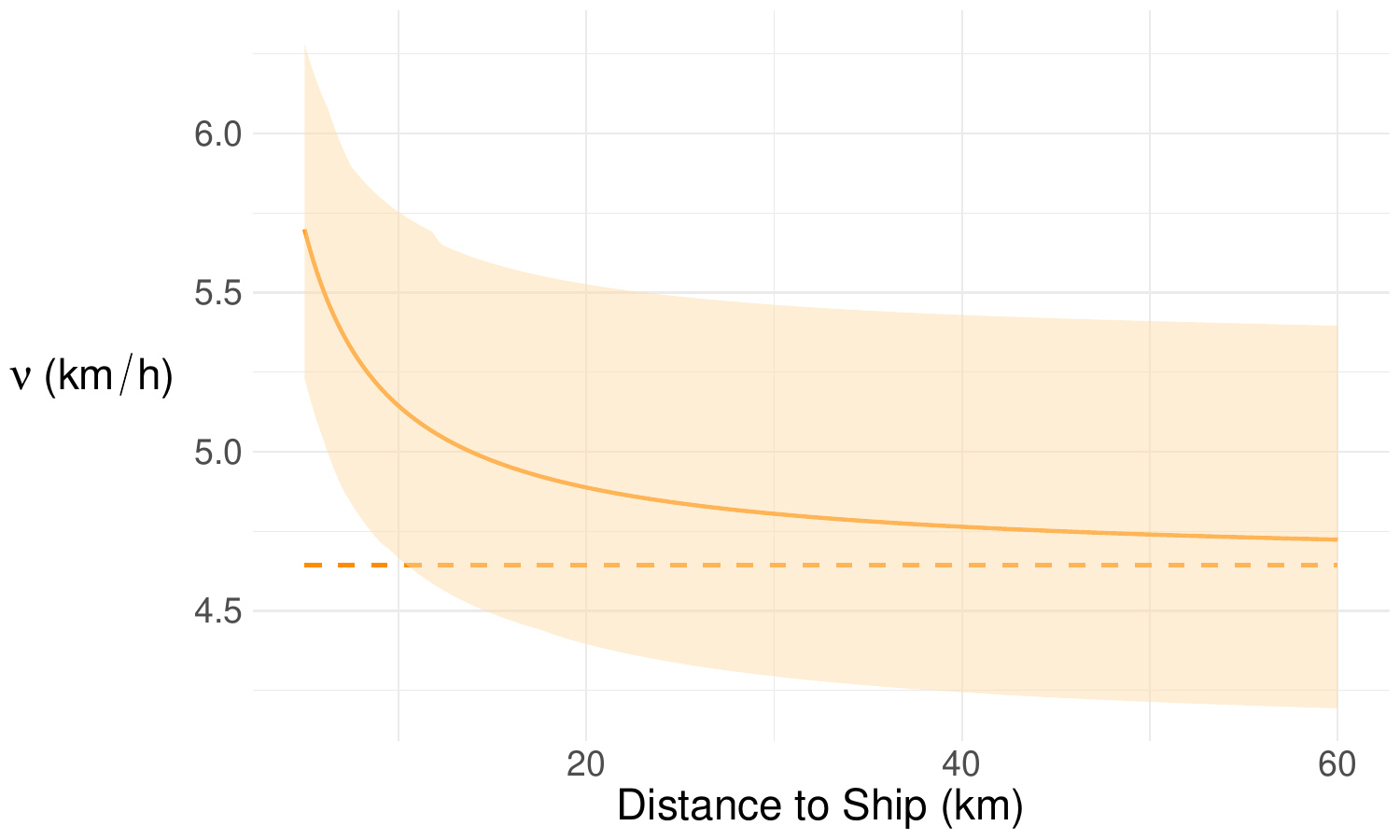"}
		\caption{}
	\end{subfigure}
	\caption{Estimated effect of ship exposure on the parameters with $95\%$ confidence intervals. The horizontal dashed lines represent the baseline values. The values on the $x$-axis are in kilometres. (a)  Population persistence parameter $\tau$. Values on the $y$-axis are in hours. (b) Population velocity parameter $\nu$. Values on the $y$-axis are in km/h.}
	\label{fig: population parameters response}
\end{figure}

The estimated $\tau$ and $\nu$ are shown in Figure~\ref{fig: population parameters response} as a function of the distance to the sound source. The value of $\nu$ increases with increased exposure to the sound. 
We draw special attention to the parameter $\tau$, which decreases with increased exposure to the ship, implying lower persistence and lower autocorrelation in the velocity of the narwhals. 
This agrees with the findings in 
\cite{heide-jorgensen_behavioral_2021} where it was observed that some narwhals react to sound exposure by changing their swimming speed and directions at distances between $5$ and $24$ km from the ship. For some individuals an increase up to  $30\%$ in the horizontal travel speed could be detected. In our model, these changes of directions and swimming speed can be interpreted as a drop in persistence due to the appearance of the ship.

The distances to the ship $D^{ship}_{\tau}(p)$ and $D^{ship}_{\nu}(p)$  for which a percentage $p$ of deviation from the population baseline values $\tau_0$ and $\nu_0$ are reached are given by 
\begin{equation}
	D^{ship}_{\tau}(p)=\frac{\alpha_{\tau}}{\log(1-p)} \quad  \mbox{and} \quad  D^{ship}_{\nu}(p)=\frac{\alpha_{\nu}}{\log(1+p)}
\end{equation}
Table~\ref{table: recovery_distances} shows these values for different proportions $p$.
For $\tau$, $90\%$ of the baseline value is recovered at distance $30.7 - (16.1,44.3)$ km. For $\nu$, $110\%$ of the baseline value is reached at distance  $11.6 - (5.2,17.6)$ km. This indicates that sound exposure can be perceived and can disturb the narwhals motion up to tens of kilometres.
Additionally, we estimate that at distance  $4.7 -(2.5,6.7)$ km from the ship, the population persistence parameter is half the baseline value. This is evidence of a strong shift in the behavior of the whales, which might have consequences on their capacity to rest and forage at short term.

Along with other studies, we believe these estimates can serve as a guideline for mitigation measures towards the effects of anthropogenic noise on the narwhals behavior. 

\renewcommand{\arraystretch}{1.75} 

\begin{table}[ht!]
	\centering
	\begin{tabular}{ |c|*{2}{w{c}{10em}|}}
		\hline
		\multirow{2}{*}{Percentage of deviation from baseline} & \multicolumn{2}{c|}{Recovery distance (km) }  \\
		\cline{2-3} 
		& $D^{ship}_{\tau}$ & $D^{ship}_{\nu}$ \\
		\hline
		$50 $ & $4.7 - [2.5,6.7] $&  $2.7 - [1.2,4.1]$ \\
        $30$ & $9.1 - [4.8,13.1]$ & $4.2 - [1.9,6.4]$ \\
		$10$ & $30.7 -[16.1,44.3]$ & $11.6 - [5.2; 17.6]$  \\
		\hline
		
	\end{tabular}
	\vspace{0.1cm}
	\caption{Estimated recovery distances of the baseline values within a $50 \%$, $30\%$ and $10\%$ range with $95\%$ confidence intervals.}
	\label{table: recovery_distances}
\end{table}

\subsection{ Baseline model checking}

\begin{figure}
\centering
    \includegraphics[scale=0.55]{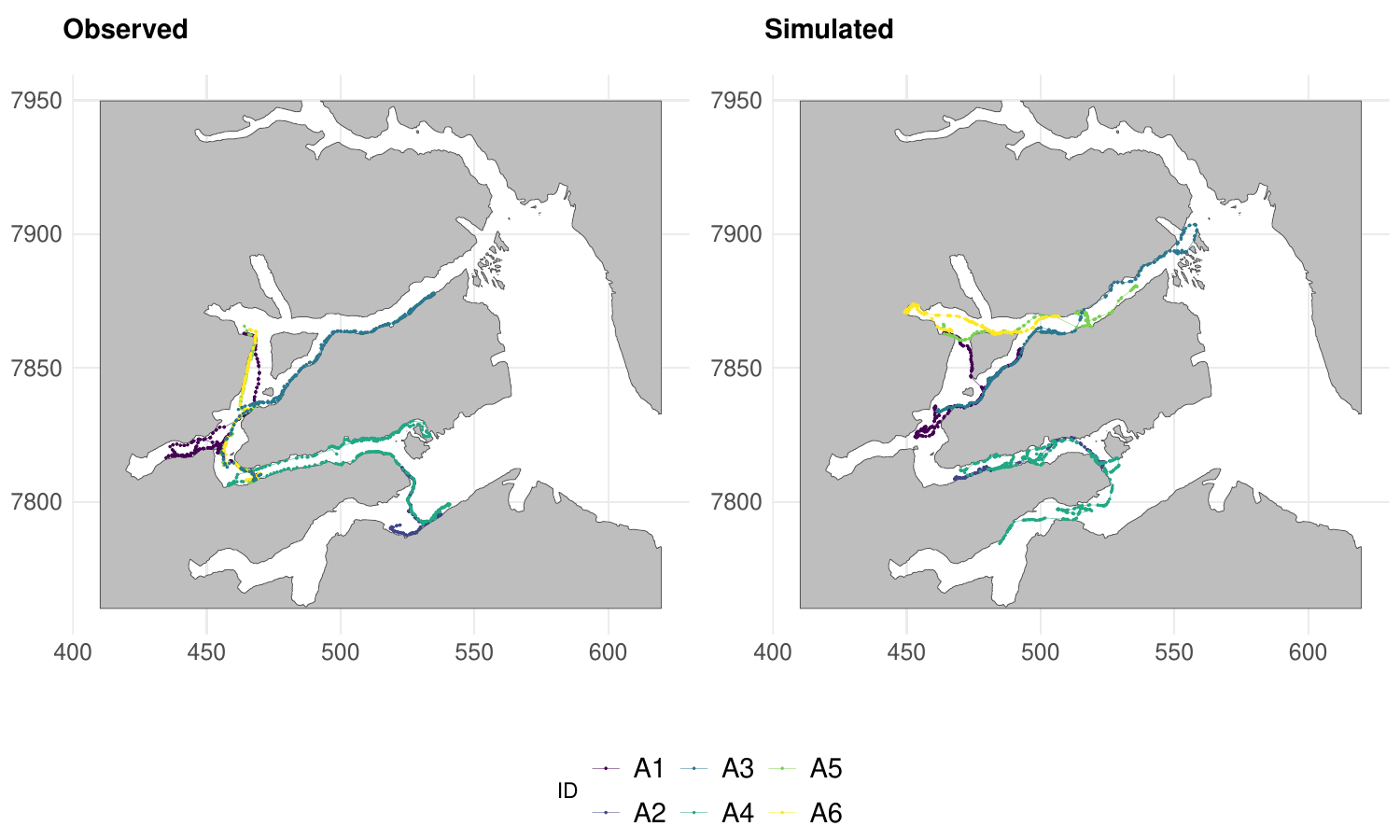}
     \caption{Observed (left) and simulated (right) trajectories. Simulated trajectories are obtained from the fitted baseline model as described in section 3.2 with a $10$ seconds time-step and then subsampled to the GPS observation times in real data.}
    \label{fig: simulation from fitted model}
   
\end{figure}

\begin{figure}[ht!]
    \centering
    \includegraphics[scale=0.4]{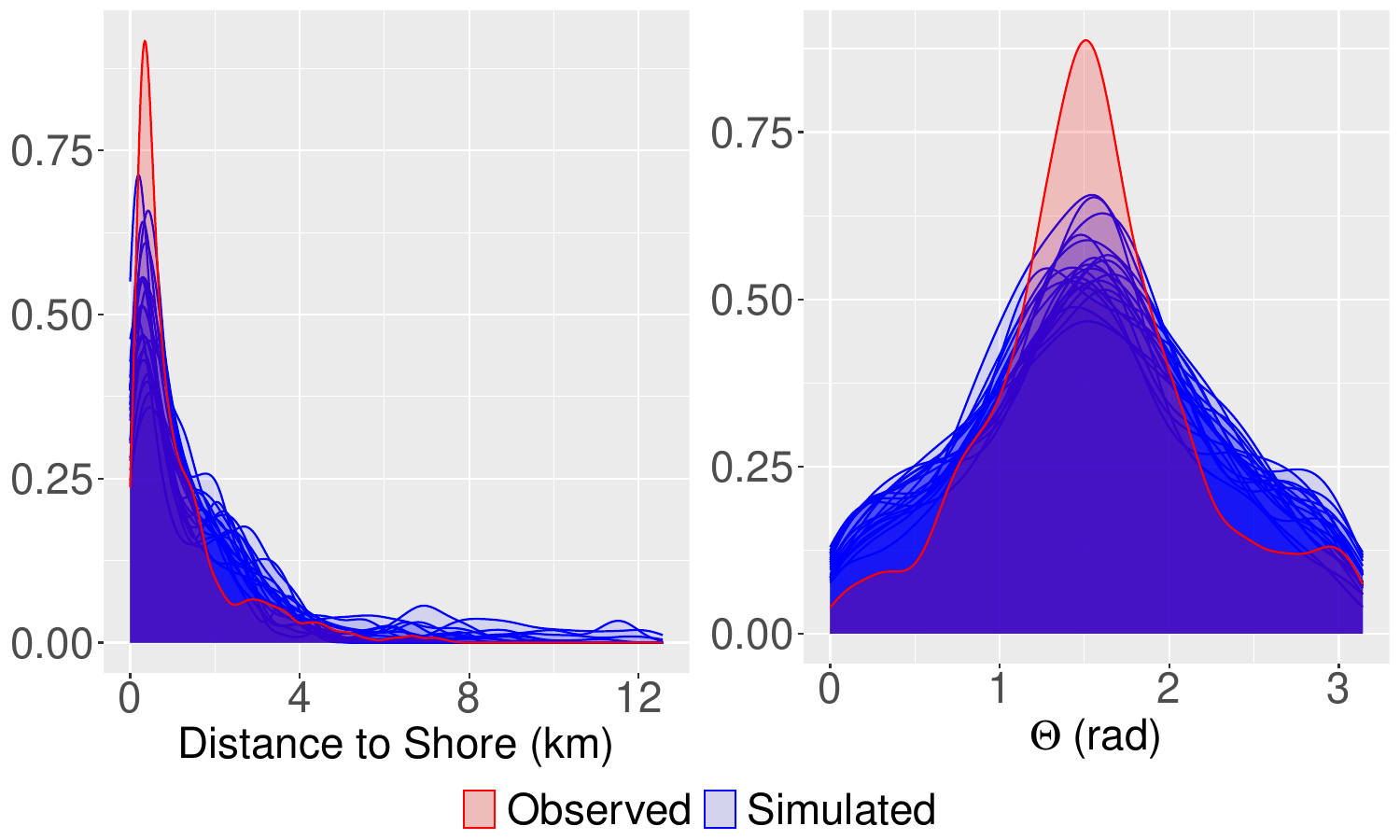}
     \caption{Densities of $D^{shore}$ and $\vert \Theta \vert $ obtained from observed and simulated trajectories.}
    \label{fig: Dshore and theta densities}
\end{figure}

We use the fitted model to simulate constrained trajectories of the six narwhals in the fjords domain, starting from the initial positions in the data with $10$-second time steps. The trajectories are then subsampled to match the observations time in the real data, and Gaussian measurement noise with a standard deviation of $50$ metres is added. Figure~\ref{fig: simulation from fitted model} illustrates simulated baseline trajectories, along with the observed trajectories before exposure. 
Next, we compute the distances $D^{shore}$ and angles $\Theta$ from the noisy trajectories. Figure~\ref{fig: Dshore and theta densities} shows the density plots obtained from $20$ simulations, alongside the density plot derived from the observed GPS data for both the angles $\Theta$ and the distances $D^{shore}$.
In terms of distance to the shore, observed and simulated data have similar characteristics, which illustrates that our model captures well the space use of the narwhals. 

%


\section{Conclusion and perspectives}


We introduced a new method to constrain a stochastic differential equation for animal movement in a bounded region of $\R^2$. Our approach relies on modeling the angular velocity as a function of the distance to the boundary and the angle between the velocity and the boundary normal vector. The additional term that constrains the motion is included in the drift, and, for this reason, acts as a confining potential.
We demonstrated how to simulate such diffusions and how to model different behaviors close to the boundary.\\

This new SDEs with smooth parameters depending on covariates was used to estimate the movement of the narwhals in the fjords. We showed that noise exposure has a significant effect on the parameters $\tau$ and $\nu$  that drive the motion by comparing a baseline and a response model. We found that Greenland narwhals movement can be affected by sound exposure over distances of up to tens of kilometers. We believe this method can be used as a basis for the assessment of behavioral responses in many contexts, and hope it will help understanding better the effects of anthropogenic noise on marine mammals movement.\\

In the future, studying theoretical properties of the SDE models we used here might be of interest. For instance, we did not exhibit clear assumptions that would guarantee that the process is effectively constrained or provide bounds on the probability to hit the boundary. We noticed that in the literature of animal movement modeling, whether that would be with SDEs, HMM or step-selection functions, spatial constraints are rarely considered. We hint that better considerations of the spatial constraints can give new insights into the interactions between marine mammals and their environment and help to understand space use.


\begin{funding}
The authors would like to thank the CNRS and IRN Madef for funding A. Delporte, and MIAI - ANR-19-P3IA-0003 for the funding of A Samson.
We also thank the project MATH-AmSud 23-MATH-12, and the UCPH-CNRS 2022 joint PhD programme.
SD was funded by Novo Nordisk Foundation, NNF20OC0062958. 
\end{funding}

\begin{supplement}
\stitle{Appendix}
\sdescription{We give more details about the narwhal data analysed in the main manuscript, derive the formulas used for estimation and illustrate how to fit our model to data in R software.}
\end{supplement}
\begin{supplement}
\stitle{Shapefile for land geometry}
\sdescription{Shapefile based on OpenStreetMap data for the land polygons in Scoresby Sound fjords system. We used high resolution satellite images to improve it with \texttt{qgis} software and applied a geometric buffer of $100$ metres.}
\end{supplement}


\bibliographystyle{imsart-nameyear} 
\bibliography{aoas-template}       
\end{document}


\begin{frontmatter}
\title{Supplementary file to "Varying coefficients correlated velocity models in complex landscapes with boundaries applied to narwhal responses to noise exposure"}
\runtitle{Movement models in complex landscapes}

\begin{aug}
\author[A]{\fnms{Alexandre}~\snm{Delporte}\ead[label=e1]{alexandre.delporte@univ-grenoble-alpes.fr}},
\author[B]{\fnms{Susanne}~\snm{Ditlevsen}\ead[label=e2]{susanne@math.ku.dk}}
\and
\author[A]{\fnms{Adeline}~\snm{Samson}\ead[label=e3]{adeline.leclercq-samson@univ-grenoble-alpes.fr}}
\address[A]{Laboratoire Jean Kuntzmann,
	Université Grenoble-Alpes\printead[presep={,\ }]{e1,e3}}

\address[B]{Department of Mathematical Sciences,
	University of Copenhagen \printead[presep={,\ }]{e2}}
\end{aug}

\begin{keyword}
\kwd{behavioral response study}
\kwd{stochastic differential equations}
\kwd{mixed effect model}
\end{keyword}

\end{frontmatter}


\appendix

\section{More details about the data}
This section is a complement to Section 2 of the paper. We provide more information about the narwhal movement data used for the analysis of behavioral disturbance.\\

\begin{figure}[ht!]
	\centering
	\includegraphics[scale=0.4]{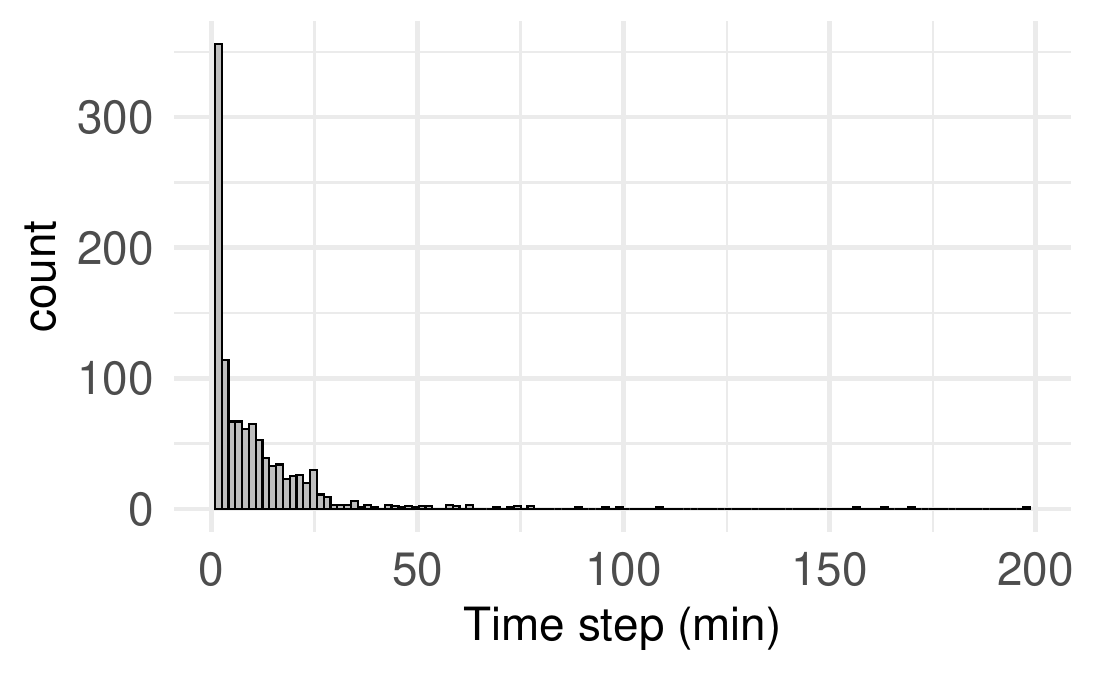}
	\caption{Histogram of time steps}
	\label{fig:alltimestepshisto}
\end{figure}

The time step between consecutive GPS observations is not constant. Its median is $4.8$ minutes and its mean is $9.3$ minutes. We show the histogram of the time steps in Figure~\ref{fig:alltimestepshisto}.

\begin{figure}[ht!]
	\centering
	\centering
	\includegraphics[scale=0.5]{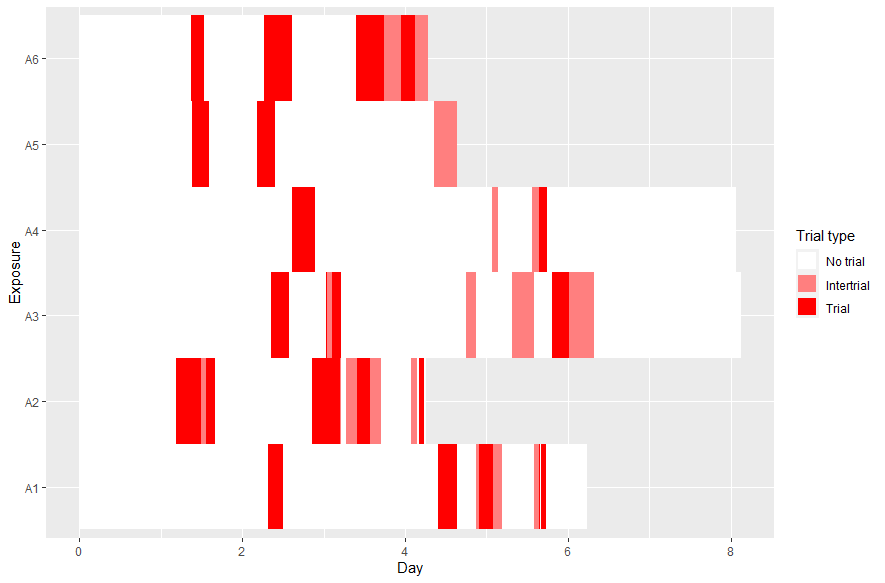}
	\caption{Trial and Intertrial periods for each narwhal}
	\label{fig: trials and intertrials distributions}
	
\end{figure}

The observations are divided into unexposed periods, for which the narwhals are not in line of sight with the ship; trial periods, when the narwhals are exposed to the ship and airguns are shot; and intertrial periods, when the narwhals are exposed to the ship but airguns are not shot. These periods are indicated by a categorical variable $T_{ship}$ in the dataset. Figure~\ref{fig: trials and intertrials distributions} shows how the exposure periods are distributed among the $6$ narwhals that were tracked. Our analysis in section 6 does not distinguish between intertrial and trial periods. They are both treated as exposure periods, though the nature and intensity of the behavioral response might differ for the two periods. We adopted this approach due to the lack of intertrial data as well as a potential persistence in time of the behavior shift due to airgun exposure during trial periods.

\begin{figure}[ht!]
	\centering
	\begin{subfigure}{0.8\linewidth}
		\centering
		\includegraphics[scale=0.5]{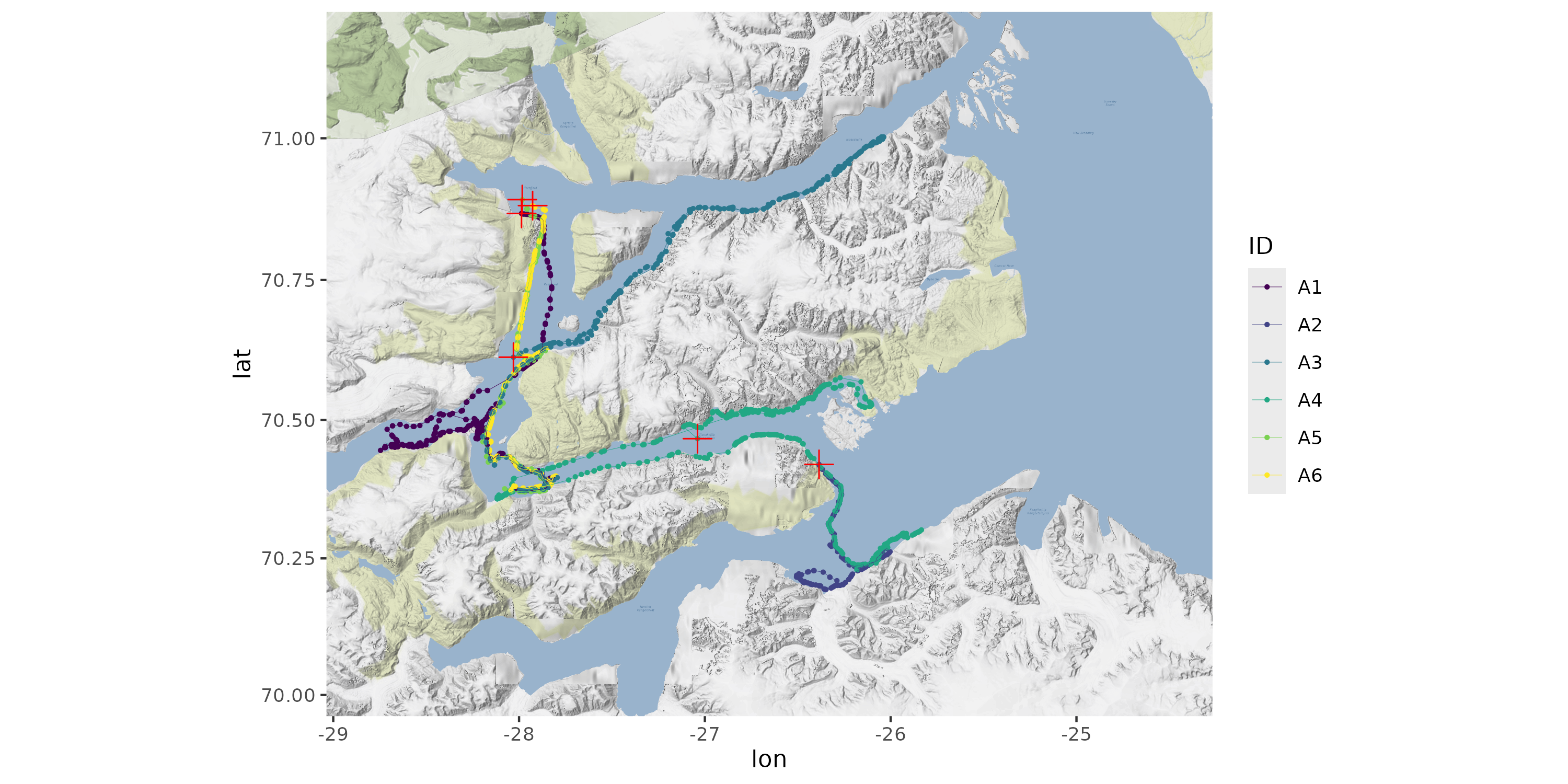}
		\caption{Tracks before exposure experiments}
	\end{subfigure}

	\begin{subfigure}{0.8\linewidth}
		\centering
		\includegraphics[scale=0.5]{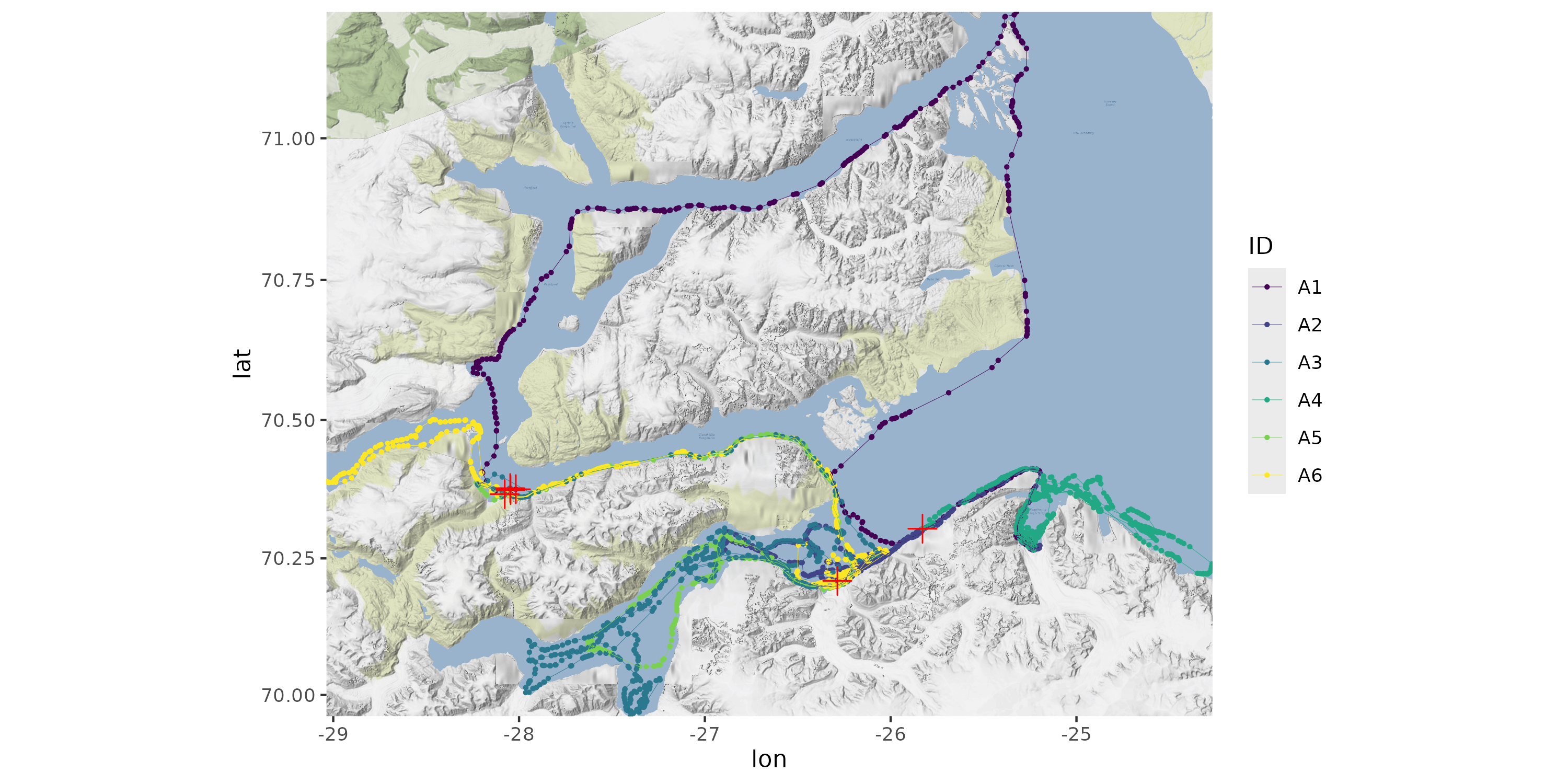}
		\caption{Tracks during exposure experiments}
	\end{subfigure}
	\caption{Movement data of East Greenland narwhals. The red crosses indicate initial positions.}
	\label{fig: tracks before and after exposure}
\end{figure}

Table~\ref{table: data distribution}  shows how the data is distributed among the different narwhals. Figure~\ref{fig: tracks before and after exposure} shows all the tracks before and after exposure with a base map of Scoresby Sound fjords system.

\begin{table}[ht!]
	\centering
	\begin{tabular}{|c|c|c|}
		\hline
		Narwhal ID & Number of measurement before exposure & Number of measurements during exposure \\
		\hline
		A1 & 354 & 576\\
		\hline
		A2  & 151 & 515 \\
		\hline
		A3 & 397 & 680 \\
		\hline
		A4 & 127 & 642  \\
		\hline
		A5 & 207 & 419\\
		\hline
		A6 & 322 & 425 \\
		\hline
		\textbf{Total} & \textbf{1558} & \textbf{3257} \\
		\hline
	\end{tabular}
	\caption{Distribution of the data among the 6 individuals}
	\label{table: data distribution}
\end{table}

All the relevant covariates   used for the analysis of narwhals movement are summarized in Table \ref{tab: summary covariates}.

\begin{table}[ht!]
	\centering
	\begin{tabular}{|c|c|p{8cm}|c|}
		\hline
		Covariate & Unit & Description & Domain \\
		\hline
		$D^{ship}(t)$ & km & distance in kilometers between the narwhal and the ship at time $t$ & $\R^+$ \\
		\hline
		$E^{ship}(t)=\frac{1}{D^{ship}(t)}$ & $km^{-1}$ & global exposure level of the narwhal to the ship disturbance at time $t$ & $\R^+$ \\
		\hline
		$D^{shore}(t)$ & km & distance between the narwhal and the nearest point on the shore at time $t$ & $\R^+$ \\
		\hline
		$I^{shore}(t)=\frac{1}{D^{shore}(t)}$ & $km^{-1}$ & global exposure level of the narwhal to the shore at time $t$ & $\R^+$ \\
		\hline
		$\Theta(t)$ & rad & angle between the vector that goes from the nearest shore point to the narwhal's position and the empirical velocity vector at time $t$  & $[-\pi,\pi]$ \\
		\hline 
	\end{tabular}
	\caption{Summary of the covariates}
	\label{tab: summary covariates}
\end{table}

\section{Proof of Proposition 3.1}
\label{section: transition density proof}

Here, we prove Proposition 3.1 The proof is inspired by the results in \citep{gurarie_correlated_2017,johnson_continuous_2008} and the proof of the transition density of the velocity process in \citep{albertsen_generalizing_2018}.
\begin{proof}
	The velocity process is an Ornstein-Uhlenbeck process. For $t \geq 0$ and $\Delta >0$,
	\begin{equation}
		V(t+\Delta)=\exp(-A\Delta) V(t)+ (I_2-\exp(-A\Delta))\mu +\frac{2\nu}{\sqrt{\pi \tau}}\int_{t}^{t+\Delta} \exp(A(s-(t+\Delta))) dW(s)
		\label{eq: RACVM solution}
	\end{equation}
	
	It has Gaussian transition density with mean 
	\begin{equation}
		\E(V(t+\Delta \vert V(t)))=\exp(-A\Delta) V(t)+ (I_2-\exp(-A\Delta))\mu 
	\end{equation}
	and covariance matrix 
	\begin{align*}Var(V(t+\Delta) \vert V(t))&=\frac{4\nu^2}{\pi \tau} \int_{t}^{t+\Delta} \exp(A(s-(t+\Delta))) \exp(A(s-(t+\Delta)))^{\top} ds\\
		&=\frac{4\nu^2}{\pi \tau} \int_{0}^{\Delta} \exp(-Au) \exp(-Au)^{\top} du 
	\end{align*}
	Since $\exp(-Au)=\exp(-\frac{u}{\tau}) R_{-\omega u}$ where $R_{-\omega u}$ is the rotation matrix with angle $-\omega u$, the matrix product  $\exp(-Au) \exp(-Au)^{\top}$ is simply $\exp(-\frac{2u}{\tau}) I_2$. We deduce that the two components of the velocity are independent and have the same variance, denoted  $q_2(\Delta)$. The variance is 
	\begin{equation}
		q_2(\Delta)=\frac{2\nu^2}{\pi}\left(1-\exp\left(-\frac{2\Delta}{\tau}\right)\right).
	\end{equation}
	These results are found in \cite{gurarie_correlated_2017}.
	In the sequel, we  use the notation 
	\[\zeta(t,s) =\frac{2\nu}{\sqrt{\pi \tau}}\int_{t}^{s} \exp(A(u-s)) dW(u).\] 
	Using that $V(s)=\mu+\exp(-A(s-t))(V(t)-\mu)+\zeta(t,s)$, we have
\begin{align*}
    X(t+\Delta) &= X(t) + \int_t^{t+\Delta} V(s) \, ds \\
    &= X(t) + \mu \Delta + \int_t^{t+\Delta} \exp(-A(s-t))(X(t) - \mu) \, ds \\
    &\quad + \int_t^{t+\Delta} \zeta(t,s) \, ds \\
    &= X(t) + \mu \Delta + \left(A^{-1}(V(t) - \mu) - A^{-1}\exp(-A\Delta)(X(t) - \mu)\right) \\
    &\quad + \int_t^{t+\Delta} \zeta(t,s) \, ds
\end{align*}
	Thus, 
	\begin{equation}
		X(t+\Delta)=X(t)+\mu \Delta+A^{-1} \left( I_2-\exp(-A\Delta)\right)(V(t)-\mu)+\xi(t,t+\Delta) 
	\end{equation}
	where $ \xi(t,t+\Delta)=\int_t^{t+\Delta} \zeta(t,s)ds$.
	The location process is also Gaussian with mean 
	\begin{equation}
		\E(X(t+\Delta) \vert V(t),X(t)) =X(t)+\mu \Delta+A^{-1} \left( I_2-\exp(-A\Delta)\right)(V(t)-\mu).
	\end{equation}
	To get an expression of the covariance matrix, first rewrite
	\begin{align*}
		\xi(t,t+\Delta)&=\int_t^{t+\Delta} \frac{2 \nu}{\sqrt{\pi \tau}} \left( \int_t^s \exp(-A((u-s))dW(u) \right) ds \\
		&= \frac{2\nu}{\sqrt{\pi \tau}} \int_t^{t+\Delta} (A^{-1}-A^{-1}\exp(A(u-t-\Delta))) dW(u) \\
		&= \frac{2\nu}{\sqrt{\pi \tau}} \int_t^{t+\Delta} A^{-1}(I_2-\exp(A(u-t-\Delta))) dW(u)
	\end{align*}
	Then use Ito's isometry
 \begin{align*}
    \mbox{Var}(X(t+\Delta)\vert X(t),V(t))&=\frac{4 \nu^2}{\pi \tau} 
    \int_{t}^{t+\Delta} A^{-1} (I_2-\exp(A(u-t-\Delta)))\\
    &\quad \times (I_2-\exp(A(u-t-\Delta)))^\top A^{-\top} \, du \\
    &=\frac{4 \nu^2}{\pi \tau} \int_{0}^{\Delta} A^{-1} (I_2-\exp(-Ar))\\
    &\quad \times (I_2-\exp(-Ar))^\top A^{-\top} \, dr.
\end{align*}
This integral can be computed explicitly since 
\[A^{-1} (I_2-\exp(-Ar))(I_2-\exp(-Ar))^\top A^{-\top}=\frac{1}{C} f(r) I_2 \]
where $f(r)=1-2\exp\left( -\frac{r}{\tau}\right)\cos(\omega r)+\exp\left(\frac{-2r}{\tau}\right)$ and $C=\frac{1}{\tau^2}+\omega^2$.
We obtain that $X_1(t+\Delta)$ and $X_2(t+\Delta)$ are independent and have the same variance, denoted $q_1(\Delta)$.
Writing $\sigma=\frac{2\nu}{\sqrt{\pi \tau}}$, the variance is
\begin{align*}
    q_1(\Delta) &= \frac{\sigma^2}{C} \left( \Delta 
    - 2 \frac{\omega \sin(\omega \Delta) - \frac{1}{\tau} \cos(\omega \Delta)}{\frac{1}{\tau^2} + \omega^2 } 
    \exp\left( -\frac{\Delta}{\tau} \right) \right. \\
    &\quad \left. + \frac{\tau}{2} \left( \frac{\omega^2 - \frac{3}{\tau^2}}{\frac{1}{\tau^2} + \omega^2} 
    - \exp\left( -\frac{2\Delta}{\tau}\right) \right) 
    \right)
\end{align*}
	
Now we compute the covariance between   $X$ and $V$ to get the full covariance matrix of $U$:
\begin{align*}
    \Gamma(\Delta) &= \frac{4\nu^2}{\pi \tau} \E\left(\left( \int_t^{t+\Delta} A^{-1}(I_2-\exp(A(u-t-\Delta))) \, dW(u)\right) \right. \\
    &\quad \left. \times \left(\int_{t}^{t+\Delta} \exp(A(s-(t+\Delta))) \, dW(s)\right)^\top\right) \\
    &= \int_t^{t+\Delta} A^{-1}(I_2-\exp(A(u-(t+\Delta)))) \exp(A(u-(t+\Delta)))^\top \, du \\
    &= \frac{4\nu^2}{\pi \tau} \int_0^{\Delta} A^{-1}(I_2-\exp(-Ar)) \exp(-Ar)^\top \, dr.
\end{align*}
	Then, 
	\[A^{-1}(I_2-\exp(-Ar)) \exp(-Ar)^\top=\frac{1}{C}\exp\left(-\frac{r}{\tau} \right) \begin{pmatrix} g(r) & h(r) \\ -h(r) & g(r) \end{pmatrix}\]
	where 
\begin{align*}
    g(r) &= \frac{1}{\tau} \left(\cos(\omega r) + \exp\left( -\frac{r}{\tau} \right)\right) 
    - \omega \sin(\omega r), \\
    h(r) &= -\frac{1}{\tau} \sin(\omega r) 
    + \omega \left(\cos(\omega r) - \exp\left( -\frac{r}{\tau}\right)\right).
\end{align*}
	Finally we get
	\[
	\gamma_1(\Delta)=\frac{\sigma^2}{2C}\left( 1+\exp\left( -\frac{2\Delta}{\tau}\right)-2\exp\left( -\frac{\Delta}{\tau}\right)-2\exp\left( -\frac{\Delta}{\tau}\right) \cos(\omega \Delta)\right),
	\]
	\[\gamma_2(\Delta)=\frac{\sigma^2}{C}\left( \exp\left( -\frac{\Delta}{\tau}\right) \sin(\omega \Delta)-\frac{\omega \tau}{2} \left(1-\exp\left( -2 \frac{\Delta}{\tau}\right) \right)\right).\]
	
	In the specific case $\omega=0$, we obtain $C=\frac{1}{\tau^2}$ and the variance of  $X$ becomes 
	\[q_1(\Delta)=\sigma^2 \tau^2\left( \Delta +2\tau\exp\left( -\frac{\Delta}{\tau}\right)+\frac{\tau}{2}\left( -3 -\exp\left( -\frac{2\Delta}{\tau}\right) \right)\right).
	\]
	Writing $\beta=\frac{1}{\tau}$ and reorganizing the terms, we obtain
	\begin{equation}q_1(\Delta)=\frac{\sigma^2}{\beta^2}\left(\Delta -2 \frac{1-\exp(-\beta \Delta)}{\beta}+\frac{1-\exp(-2\beta \Delta)}{2\beta}\right).
	\end{equation}
	This result match equation $(6)$ in \citep{johnson_continuous_2008}. Similarly, in the case $\omega=0$, we get $\gamma_2=0$ and the expression for $\gamma_1$ match equation $(7)$ in \citep{johnson_continuous_2008}.
	
\end{proof}

\section{Measurement error}

In the application in Section 6, different values of the measurement error were estimated for the data before and after exposure: $35$ m before and $48$ m after. Both these values are consistent with the results in \citep{wensveen_path_2015}. However, the post exposure estimation gave non-positive definite Hessian matrix for the negative log-likelihood, which prevents from using the information matrix equality to get confidence intervals of the estimates. Fixing a $35$ m measurement error value when fitting the response model led to the same issue. We therefore tried different values of the measurement error, and kept the one that gave a positive definite hessian matrix and had the highest log-likelihood value. It turned out to be $50$ m, very close to the initially estimated $48$ m. Table~\ref{tab: estimates for fixed measurement errors} shows these results. In comparison, the final values of the log-likelihood when $\sigma_{obs}$ is estimated from the data are respectively $4273$ and $8043$ before and after exposure, while the estimate of $\tau_0$ is $1.10$, and the estimates of $\alpha_\tau$ and $\alpha_\nu$ are respectively $-4.19$ and $0.66$, which is in the confidence interval of the final estimations we kept (those obtained for $\sigma_{obs}=50$ m).
\begin{table}[ht!]
	\centering
	\begin{tabular}{|c|c|c|c|c|c|c|}
		\hline
		$\sigma_{obs}$ (m) & $\hat{\tau_0}$ & Baseline llk & Response llk & P.d hessian & $\alpha_{\tau}$ & $\alpha_\nu$ \\
		\hline
		$30$  & $0.96 \pm 0.15$ & $4261$  & 8038 & No & $0.29$ & $2.17$ \\
		\hline
		$40$  & $1.18 \pm 0.16$   & $4266$ & 8014  & No  & $-2.06$ & $0.60$\\
		\hline
		$45$  & $1.29 \pm 0.17$   & $4243$ & 7965  & No & $-3.64$ & $0.49$\\
		\hline
		$\mathbf{50}$  & $\mathbf{1.35 \pm 0.16}$ & $\mathbf{4208}$ & $\mathbf{7861}$ & \textbf{Yes} & $\mathbf{-3.43 \pm 0.70}$ & $\mathbf{0.74 \pm 0.27}$\\
		\hline
		$75$ &  $1.63 \pm 0.19$ & $3944$ & $7225$ & Yes & $-3.88 \pm 0.73$ & $0.76 \pm 0.30$\\
		\hline 
		$100$ & $1.85 \pm 0.22$ & $3640$ & $6590$ & Yes & $-4.27 \pm 0.74$ & $0.63 \pm 0.29$ \\
		\hline 
	\end{tabular}
	\caption{Estimate for the baseline and response models for several fixed measurement errors.}
	\label{tab: estimates for fixed measurement errors}
\end{table}

\section{Code example}

We illustrate briefly how to fit our baseline SDE model and obtain the results with \texttt{smoothSDE} R package \cite{michelot_varying-coefficient_2021}. The version of the package including our new model is available \href{https://github.com/alexandre-delporte/smoothSDE}{here} . We suppose the package has been loaded. We consider a dataframe \texttt{dataBE} containing the preprocessed observations before exposure to the ship in the columns \texttt{x} and \texttt{y}, an animal identifier in a column \texttt{ID} and columns \texttt{Ishore} and \texttt{AngleNormal} for the covariates $I^{shore}$ and $\Theta$. The first step consists in choosing initial SDE parameters and model formulas. For the model we consider, there are five parameters $\mu_1$, $\mu_2$, $\tau$, $\nu$, and $\omega$, and each of them needs a formula. Specification of the formulas is identical to the \texttt{R} package \texttt{mgcv}. Among the parameters, $\mu_1$, $\mu_2$ will be set to 0, while $\tau$, $\nu$ and $\omega$ are expressed as in section 4.1.

\begin{lstlisting}[language=R]
#number of observation
n_pre<-nrow(dataBE)

#initial parameters
par0 <- c(0,0,1,4,0)

#model formulas
formulas <- list(mu1 = ~1 ,mu2 =~1,tau =~s(ID,bs="re"),nu=~s(ID,bs="re"),
omega=~ti(AngleNormal,k=5,bs="cs")+ti(Ishore,k=5,bs="cs")+
ti(AngleNormal,Ishore,k=c(5,5),bs="cs"))
\end{lstlisting}

We then specify the measurement error for each observation in an array of covariance matrices. We suppose they are all diagonal with the same standard deviation \texttt{sigma\_obs}. We will fix this measurement error. \\

\begin{lstlisting}[language=R]
# 50m measurement error
sigma_obs=0.05
H=array(rep(sigma_obs^2*diag(2),n_pre),dim=c(2,2,n_pre))
\end{lstlisting}

We can then create the SDE object as in \cite{michelot_varying-coefficient_2021}. We choose the type of SDE in the argument \texttt{type}. Here, it is \texttt{RACVM} (see Section 3.1) since we want to include a non zero rotation parameter $\omega$. The name of the columns where the observations are found is specified in the \texttt{response} argument.
We specify the measurement error matrix \texttt{H} in the argument \texttt{other\_data}. Fixed parameters are indicated in the argument \texttt{fixpar}.

\begin{lstlisting}[language=R]
#create SDE object
baseline_50m<- SDE$new(formulas = formulas,data = dataBE,type = "RACVM",
response = c("x","y"),par0 = par0,other_data=list("H"=H),
fixpar=c("mu1","mu2"))
\end{lstlisting}

To fix specific parameters in the statistical model, we need to use the \texttt{map} attribute. Here we use it to specify that the smoothing parameters should be fixed. Then we update the smoothing parameters to $1$, and fit the SDE model.

\begin{lstlisting}[language=R]
#update map to fix smoothig parameters
baseline_50m$update_map(list("log_lambda"=factor(c(1,2,rep(NA,4))))

#update smoothing parameters values
init_lambda=rep(1,6)
baseline_50m$update_lambda(init_lambda)

#fit the model
baseline_50m$fit()
\end{lstlisting}

The results of the optimization are stored in the attribute \texttt{tmb\_rep}.
We can extract the estimated parameters along with the standard errors.

\begin{lstlisting}[language=R]
#estimates
estimates_bas_50m=as.list(baseline_50m$tmb_rep(),what="Est")
#standard error
std_bas_50m=as.list(baseline_50m$tmb_rep(),what="Std")
\end{lstlisting}

Finally, we would like to plot all the smooth parameters as a function of the covariates. 
We can do it with the \texttt{get\_all\_plots} method.
We only need to specify the range of each covariate value we want to plot, a link function if we don't want to have directly the covariate on the x-axis but rather a function of the covariate, and the x-axis label of the plots. We put the option \texttt{show\_CI="pointwise"} to show the pointwise confidence intervals on the plots.

\begin{lstlisting}[language=R]
#range of the covariates
D_low=0.073
D_up=3
xmin=list("Ishore"=1/D_up)
xmax=list("Ishore"=1/D_low)
#link function
link=list("Ishore"=(\(x) 1/x))
#label
xlabel=list("Ishore"="Distance to shore")

#draw plots
plots_bas_50m=baseline_50m$get_all_plots(model_name="baseline_50m",
xmin=xmin,xmax=xmax,link=link,xlabel=xlabel,show_CI="pointwise",save=TRUE)
\end{lstlisting}


\bibliographystyle{imsart-nameyear} 
\bibliography{biblio-supp}       